\documentclass[iop,apj,tighten]{emulateapj}
\usepackage{apjfonts}

\usepackage{bm}
\usepackage{amsmath,amstext}
\usepackage[breaklinks,colorlinks,citecolor=blue,linkcolor=magenta]{hyperref}
\usepackage[all]{hypcap} 

\shortauthors{Hezaveh et al.}
\shorttitle{Substructure in SDP.81}

\usepackage{color}

\newcommand{\CN}{\ensuremath{{\bf C}_N}}
\newcommand{\Cs}{\ensuremath{{\bf C}_s}}
\newcommand{\B}{\ensuremath{{\bm{W}}}}
\newcommand{\A}{\ensuremath{{\bf R}}}
\newcommand{\Cp}{\ensuremath{{\bf C}_{\rm p}}}
\renewcommand{\L}{\ensuremath{{\cal L}}}
\newcommand{\T}{{\sf T}}

\newcommand{\dOdp}{\frac{\partial \bm{M}}{\partial \bm{p}}}

\newcommand{\psub}{\ensuremath{\bm{p}_{\rm sub}}}

\newcommand{\LOGP}{\ensuremath{{\cal  E} }}
\newcommand{\Aa}{\ensuremath{{\bf A}}}
\newcommand{\Bb}{\ensuremath{{\bf B}}}

\begin{document}

\title{Detection of lensing substructure using ALMA observations of the dusty galaxy SDP.81}

\author{Yashar~D.~Hezaveh\altaffilmark{1,12},
Neal~Dalal\altaffilmark{2,3,4,5}, 
Daniel~P.~Marrone\altaffilmark{6},   
Yao-Yuan~Mao\altaffilmark{1,7},
Warren~Morningstar\altaffilmark{1}, 
Di~Wen\altaffilmark{2},
Roger~D.~Blandford\altaffilmark{1,7},
John~E.~Carlstrom\altaffilmark{8}, 
Christopher~D.~Fassnacht\altaffilmark{9},
Gilbert~P.~Holder\altaffilmark{10}, 
Athol~Kemball\altaffilmark{2},
Philip~J.~Marshall\altaffilmark{7},
Norman~Murray\altaffilmark{11,13},
Laurence~Perreault~Levasseur\altaffilmark{1},
Joaquin~D.~Vieira\altaffilmark{2},
Risa~H.~Wechsler\altaffilmark{1,7}
}

\altaffiltext{1}{Kavli Institute for Particle Astrophysics and Cosmology and Department of Physics, Stanford University, 452 Lomita Mall, Stanford, CA 94305-4085, USA}
\altaffiltext{2}{Astronomy Department, University of Illinois at Urbana-Champaign, 1002 W.\ Green Street, Urbana IL 61801, USA}
\altaffiltext{3}{School of Natural Sciences, Institute for Advanced Study, 1 Einstein Drive, Princeton, NJ 08540, USA}
\altaffiltext{4}{Kavli Institute for the Physics and Mathematics of the Universe, TODIAS, The University of Tokyo, Chiba, 277-8583, Japan}
\altaffiltext{5}{Department of Chemistry and Physics, University of Kwa-Zulu Natal, University Road, Westville, KZN, South Africa}
\altaffiltext{6}{Steward Observatory, University of Arizona, 933 North Cherry Avenue, Tucson, AZ 85721, USA}
\altaffiltext{7}{Kavli Institute for Particle Astrophysics and Cosmology and Department of Particle Physics and Astrophysics; SLAC National Accelerator Laboratory, Menlo Park, CA 94305, USA}
\altaffiltext{8}{Kavli Institute for Cosmological Physics, University of Chicago, 5640 South Ellis Avenue, Chicago, IL 60637, USA}
\altaffiltext{9}{Department of Physics, University of California, One Shields Avenue, Davis, CA 95616, USA}
\altaffiltext{10}{Department of Physics, McGill University, 3600 Rue University, Montreal, Quebec H3A 2T8, Canada}
\altaffiltext{11}{CITA, University of Toronto, 60 St.\ George St., Toronto ON M5S 3H8, Canada}
\altaffiltext{12}{Hubble Fellow}
\altaffiltext{13}{Canada Research Chair in Astrophysics}

\begin{abstract}
\noindent
We study the abundance of substructure in the matter density near galaxies using ALMA Science Verification observations of the strong lensing system SDP.81. We present a method to measure the abundance of subhalos around galaxies using interferometric observations of gravitational lenses.  Using simulated ALMA observations, we explore the effects of various systematics, including antenna phase errors and source priors, and show how such errors may be measured or marginalized.  We apply our formalism to ALMA observations of SDP.81.  We find evidence for the presence of a $M=10^{8.96\pm 0.12} M_{\odot}$  subhalo near one of the images, with a significance of $6.9\sigma$ in a joint fit to data from bands 6 and 7; the effect of the subhalo is also detected in both bands individually. We also derive constraints on the abundance of dark matter subhalos down to $M\sim 2\times 10^7 M_{\odot}$, pushing down to the mass regime of the smallest detected satellites in the Local Group, where there are significant discrepancies between the observed population of luminous galaxies and predicted dark matter subhalos.  We find hints of additional substructure, warranting further study using the full SDP.81 dataset (including, for example, the spectroscopic imaging of the lensed carbon monoxide emission).  We compare the results of this search to the predictions of $\Lambda$CDM halos, and find that given current uncertainties in the host halo properties of SDP.81, our measurements of  substructure are consistent with theoretical expectations.  Observations of larger samples of gravitational lenses with ALMA should be able to improve the constraints on the abundance of galactic substructure.
\end{abstract}

\keywords{gravitational lensing: strong --- dark matter}

\section{Introduction}

Over the past few decades, multiple probes of the large-scale structure of the universe have established a clear picture of our cosmology that is described well by a simple model, the $\Lambda$CDM model \citep[e.g.,][]{Dodelson03}.  This model predicts that structures including galaxies arise through a process of hierarchical structure formation, driven by gravitational instability of cold dark matter (CDM) density perturbations seeded by quantum fluctuations generated during an early epoch of exponential expansion called inflation.  On large scales ($\gtrsim$ megaparsecs) and across all epochs, the predictions of this model agree remarkably well with a wide variety of observations, including cosmic microwave background (CMB) anisotropies \citep{Planck2015}, galaxy clustering \citep{BOSS2014}, and weak gravitational lensing \citep{CFHTLens}.  On smaller, sub-galactic scales where structure has become highly nonlinear (e.g., $\lesssim$ 10 kpc) at low redshift, the agreement between CDM predictions and observations of galactic structure has been less clear.

One of the most famous examples of CDM's small scale difficulties is the Missing Satellite Problem \citep[e.g.,][and references therein]{kravtsov:10}. Numerical simulations of the formation of galactic dark matter halos robustly predict the presence of a large population of bound subhalos orbiting within all virialized objects such as the Milky Way.  These subhalos span a broad spectrum of masses, from objects like the Magellanic Clouds, extending down to the resolution limits of the simulations $M\sim 10^3 M_\odot$ \citep{ViaLactea2,GHalo,Aquarius}.
Observations of dwarf galaxies in the Local Group find an almost factor of 10 deficit of satellites compared to $\Lambda$CDM predictions, at masses corresponding to $V_{\rm circ} \sim 10\,$km/s \citep{kravtsov:10}.

Two main classes of solutions have been proposed to resolve this discrepancy.  The first solution involves modifying the microphysics of dark matter particles, in a manner that would suppress the abundance of low mass substructure while remaining consistent with existing bounds on structure at $k \lesssim 10 \,h\, {\rm Mpc}^{-1}$ from the  Lyman $\alpha$ forest \citep{seljak:06,markovic:14}.  Examples of proposed modifications to CDM include warm dark matter (WDM), in which the thermal streaming motions of dark matter (DM) particles wipe out small scale structure \citep{Bode01,Abazajian06}, or self-interacting dark matter, in which DM particles can have non-gravitational interactions \citep{Spergel00}.  In principle, fluid descriptions of dark matter could also suppress satellites \citep{khoury:15}.

Alternatively, the solution to the Missing Satellite Problem may lie within baryonic astrophysics.  Numerous processes have been suggested for suppressing the efficiency of star formation within low mass halos and subhalos, such as photoevaporation during reionization, or feedback from massive stars, supernovae or black holes \citep[see][for a review]{kravtsov:10}.
In this scenario, a large number of dark matter subhalos should exist around most galaxies, however most of them would be devoid of stars and therefore difficult to detect using conventional means.

Gravitational lensing has long been suggested as a method of distinguishing between these possibilities \citep[various properties of tidal streams and their disruption have also been suggested as a method to detect subhalos in the Milky Way, e.g.,][]{Siegal-Gaskins:08, Carlberg:09, Erkal:15a,Erkal:15b}.  Strong gravitational lensing is sensitive to galactic substructure, even if that substructure is completely devoid of baryons \citep[e.g.,][]{Mao:98,Metcalf01,dalal:02,Moustakas03,KochDalal04,Koopmans:05, Keeton06,Vegetti:09,vegetti:12,Nierenberg:14,Xu:15A,Cyr-Racine:15}.  \cite{dalal:02} analyzed a sample of strongly lensed quasars and found that the lenses must have a large abundance of substructure: they found that the fraction of projected mass within $\sim 10\,$kpc of lens galaxies in the form of substructure was $0.3\% < f_{\rm sub} < 3.5\%$, at 90\% confidence.  \cite{vegetti:14} analyzed a sample of strongly lensed galaxies observed using Hubble Space Telescope (HST) and Keck AO data and found  $0.2\% < f_{\rm sub} < 2.8\%$, at 68\% confidence, with two detections reported in \citet{vegetti:10} and \citet{vegetti:12}.   These measurements of projected substructure are in broad agreement with expectations for the  massive elliptical galaxies that are typical of strong lenses (see Section~\ref{sec:lcdm}), but the large uncertainties do not yet allow for a discriminating test of dark matter models.

Recently, \citet{Hezaveh13A} proposed that ALMA observations of the large population of strongly lensed dusty galaxies discovered in sub-mm surveys like the South Pole Telescope \citep{vieira:13,hezaveh:13b}, Atacama Cosmology Telescope \citep{marsden:14} or Herschel \citep[][]{negrello:10,wardlow:13, Bussmann:13} could significantly improve constraints on dark matter substructure.  Towards that end, in this paper we present a method for detecting dark matter subhalos using interferometric data, show a few examples of the performance of the pipeline on mock data and finally, we apply our formalism to the recent ALMA Science Verification observations of the strongly lensed dusty galaxy SDP.81 \citep{SDP81}.

The outline of the paper is as follows.
In Section~\ref{sec:framework}, we present a formalism for modeling extended lensed sources using interferometric observations. In Section~\ref{sec:subhalofinding_method} we present a perturbative method for searching for subhalos using interferometric observations. Section~\ref{sec:implement} describes our implementation of this method and presents the results of various tests and analysis of mock data used to test the pipeline. 
In Section~\ref{sec:results} we analyze ALMA observations of SDP.81 using this pipeline, quantifying the abundance of subhalos in the lensing galaxy.
In Section~\ref{sec:lcdm} we compare the constraints obtained in the previous section to the predictions of the $\Lambda$CDM model and finally discuss the results in Section~\ref{sec:discuss}.
For all of the modeling presented in this paper, we assume a flat $\Lambda$CDM expansion history with mean matter density $\Omega_M = 0.267$ and current expansion rate $H_0 = 71$ km Mpc$^{-1}$ s$^{-1}$.

\section{Framework: fitting visibilities with a pixelated source}
\label{sec:framework}

Data from interferometric observations consist of a large number of complex visibilities.  
We fit these data using a model for the distribution of matter in the lensing galaxy, the background source emission, and certain aspects of the measurements such as time-varying antenna phase errors.  To describe the source emission, we use a pixelated source map containing many pixels for each observed channel. We can think of the source map pixels as parameters in our lens model, along with parameters describing the lens mass distribution, and other nuisance parameters like antenna phase errors.  Below, we will use the notation $\bm{\eta}$ to denote lens mass parameters (see Table~\ref{lens_pars}), $\bm{S}$ to denote source pixel parameters, $\bm{\phi}$ to denote other parameters like phase errors, and $\bm{p}$ to denote the full set of parameters, i.e., $\bm{p} = \{\bm{\eta}, \bm{S}, \bm{\phi} \}$.

The general framework of fitting strongly lensed images with pixelated sources is described in detail in \citet{warren:03} and \citet{suyu:06}.  In general, given data vector $\bm{D}$, model predictions $\bm{M}(\bm{p})$, which depend on lensing parameters $\bm{\eta}$ (e.g., lensing mass) and source parameters $\bm{S}$ (i.e., pixel values), we can write  the posterior probability distribution (PDF) for the parameters as ${\cal L} \propto e^{-\LOGP/2}$, where \LOGP \ is defined as
\begin{align}
\label{chi2}
\LOGP(\bm{p}) &=  (\bm{D} - \bm{M}(\bm{p}))^\T \CN^{-1} (\bm{D} - \bm{M}(\bm{p})) + \bm{p}^\T \Cp^{-1} \bm{p} \\
 & = (\bm{D} - \bm{M})^\T \CN^{-1} (\bm{D} - \bm{M}) 
+ \bm{S}^\T \Cs^{-1} \bm{S} + \bm{\phi}^\T {\bf C}_\phi^{-1} \bm{\phi}. \nonumber
\end{align}
In this expression, the first term on the right hand side corresponds to the usual goodness-of-fit $\chi^2$, and the second term corresponds to a prior on the model parameters $\bm{p}$.  
$\CN$ is the noise covariance matrix and $\Cp$ is the prior covariance matrix describing our assumed multi-variate Gaussian prior PDF for source parameters.  We assume that the covariance matrix is block diagonal, $\Cp = {\rm diag}({\bf C}_\eta, \Cs, {\bf C}_\phi)$
i.e., we assume no prior covariance between lens parameters, source parameters, and phase parameters.  In the second line of Equation~\eqref{chi2}, and for the rest of this paper, we set ${\bf C}_\eta^{-1} = 0$, i.e., we assume no prior on the lens model parameters.

Of particular importance is the prior on source pixel parameters $\bm{S}$.  We denote that block within \Cp\ as \Cs,  the source prior covariance matrix, which gives a term $\bm{S}^\T\Cs^{-1}\bm{S}$ in Equation~\eqref{chi2}.  This matrix is often written as $\Cs=\lambda \hat{\Cs}$, where $\lambda$ is a scaling parameter \citep{suyu:06}. We describe our procedure for determining the strength of the source prior, $\lambda$, in Section~\ref{sec:prior}.
In Equation~\eqref{chi2}, the parameter vector $\bm{S}$ is defined without loss of generality so that the Gaussian prior is centered at $\bm{S}=0$.  More generally, if we define $\bm{S}$ so that the prior is centered at some $\bm{S}_c$, then we replace $\bm{S} \to \bm{S}-\bm{S}_c$ in the second to last term in Equation~\eqref{chi2}.

We note that the number of source parameters $\bm{S}$ is quite large, and as noted in previous works \citep[e.g.,][]{suyu:06}, the source prior will act to regularize the reconstruction of the source parameters and avoid over-fitting the data. 
This Gaussian prior, described by covariance matrix \Cs,  is discussed below in Section~\ref{sec:prior}. 
For visibility data, the noise covariance \CN\ is diagonal, and its amplitude can be determined from the data, using a method described in detail in Section~\ref{sec:implement}. 

For ALMA observations, the data vector $\bm{D}$ consists of the real and imaginary parts of complex visibilities. We can write our model visibilities as
\begin{equation}
\bm{M}({\bm {\eta,S,\phi}}) = {\bf F}(\bm{\phi})\, {\bf B \, L({\bm \eta})\,} \bm{S}\,,
\label{linearsource}
\end{equation}
where $\bm{S}$ is a vector of source pixel values, ${\bf L}$ is a lensing operator that maps the brightness of each source pixel to the image pixels (sky emission),  ${\bf B}$ is a diagonal primary beam operator that multiplies the sky emission with the primary beam, and ${\bf F}$ is a dense Fourier operator, whose $ij$-th element is equal to $e^{-i (\bm{k}_i\cdot\bm{r}_j + \phi_{l1}-\phi_{l2})}$, corresponding to a visibility at $uv$-coordinate $\bm{k}_i$ from baseline $l$ (composed of two antennas, labeled $l1$ and $l2$), and an image pixel located at $\bm{r}_j$. Note that rows of ${\bf F}$ with equal values of $l$ have a common phase error (e.g., visibilities from the same baseline within the segmentation time of the antenna phase corruption parameters). 
To calculate ${\bf L}$ for a set of lens parameters ${\bm \eta}$ (e.g., mass and ellipticity) we solve the non-linear lensing equation using a ray-tracing approach. 
Note that ${\bf L}$, ${\bf B}$, and ${\bf F}$ are all linear operators, and that $\bm{S}$ is a subset of the model parameter vector $\bm{p}=\{\bm{\eta , S, \phi}\}$.   
Application of ${\bf F\, B}$ to a sky emission model produces $\bm{M}$, the vector of the model visibilities.

We treat the source parameters ${\bm S}$ (the source pixel intensities) as nuisance parameters, and marginalize over them.  
Our goal is then to compute the lensing parameter posterior described by Equation~\eqref{chi2}, marginalized over source parameters.
Because the observables are linear in the source pixels, and \LOGP \  is quadratic in the observables, then the likelihood is a Gaussian function of the source pixels.  Since our assumed source prior is also Gaussian, the posterior is Gaussian.  This allows us to analytically marginalize over the source nuisance parameters using Gaussian integrals, to determine the posterior PDF for the remaining parameters.  

The difference in marginalized log-posterior between two models is then
\begin{equation}
\Delta \LOGP = \Delta \left[ \Bb^{\T} \Aa^{-1} \Bb - \log(\det[\Aa]) \right]\,,
\end{equation}
where $\Aa = ({\bf FBL})^\T \CN^{-1} ({\bf FBL})+\Cs^{-1}$ and $\Bb~=~{\bf (L B F)}^{\T} \CN^{-1} {\bm D}$ \citep{Suyu:10}. 
The source reconstruction that maximizes the unmarginalized posterior (at fixed lens model) is also analytic, and is given by \citep{warren:03,suyu:06}
\begin{equation}
\bm{S} = {\Aa}^{-1} \Bb.
\label{reconstruct}
\end{equation}

We note that the above formalism is general to multi-frequency data. Vectors $\bm{D}$ and $\bm{S}$ can be concatenations of multiple data and reconstructed source vectors in different frequencies, while matrices \CN \ and \Cs \ could be formed as block diagonal matrices including the noise and source prior in each channel. It is also possible to include regularization between different frequencies by allowing non-zero elements in \Cs \ in off-block-diagonal elements.

\subsection{Source Structural Prior}
\label{sec:prior}

As Equation~\eqref{chi2} makes explicit, the posterior PDF depends on our choice of the source prior, \Cs.  Various forms of source priors are explored in the literature, including so-called gradient and curvature priors \citep{warren:03,suyu:06}. More generally, we could employ more physically motivated priors that are based on the expected morphology of the specific sources under investigation.

The dust and molecular line morphologies of early star forming galaxies are expected to be well represented by a number of star forming clumps embedded in a larger structure, such as an exponential disk.  One can  use this structure to construct a source prior by calculating the power spectrum and covariance of such a clustered source model. Suppose that we have $N_c$ clumps in our source galaxy, whose distribution within the galaxy has profile $U_c(r)$.  We normalize $U_c$ to have unit integral, $\int U_c (r) d^2r =1$.  Its Fourier transform is $U_c(k)$.  Clump $i$ has luminosity $L_i$ and profile $u_i(r)$, normalized to have unit integral, $\int u_i(r)d^2r = 1$. Then the power spectrum of the source emission is proportional to 

\begin{equation}
P_{\rm src}(k) \propto \left[ \sum_i^{N_c}  L^2_i |u_i(k)|^2 + \sum_{i\neq j}^{Nc} L_iL_j |U_c(k)|^2 u^{*}_i(k) u_j (k) \right]\,.
\label{sourcePk}
 \end{equation}  

The Fourier transform of this power spectrum gives the correlation function of the source emission, and the source covariance \Cs\ is determined from this correlation function.  Note that the normalization of the power spectrum (and hence the normalization of \Cs) has not been specified.  In principle, we could normalize \Cs\  using the observed intensity, but this has been magnified by an (a priori) unknown lensing magnification.  

Instead, we normalize \Cs\ by maximizing the marginalized likelihood (Bayesian evidence) for the fixed-parameter lens model, as discussed in \citet{suyu:06}.  We can scale an arbitrarily normalized source covariance matrix $\hat{\Cs}$ (which in essence only defines the form of the prior), to get an appropriately scaled matrix $\Cs=\lambda \hat{\Cs}$, where $\lambda$ is a scaling parameter which can be determined by solving
\begin{equation}
N_s - \lambda \mathrm{Tr}([{\bf FL} +  \hat{\Cs} ]^{-1}\hat{\Cs}^{-1}) - \lambda \bm{S}^\T  \hat{\Cs}\bm{S} = 0\,,
\label{lambda_norm}
 \end{equation}  
where $N_s$ is the number of source pixels and $\bm{S}$ is determined using Equation~\eqref{reconstruct}. This equation can then be solved non-linearly.

\begin{figure*}
\begin{center}
\centering
\includegraphics[trim= 0 0 0 0, clip, width=0.90\textwidth]{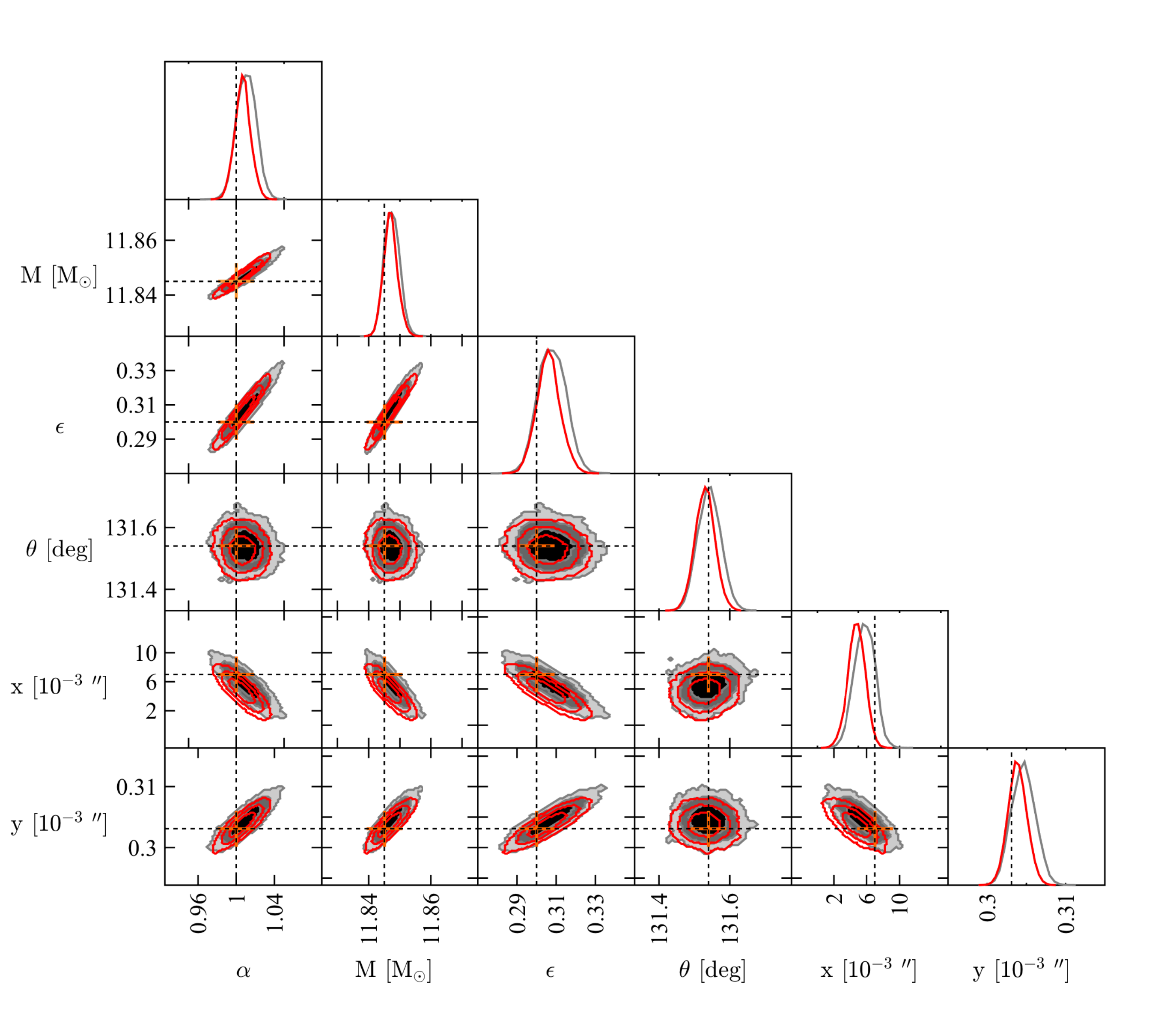}
\centering
\end{center}
\caption{Posterior PDF for the lens model parameters for a mock dataset of a 2-hour long observation of a 50 mJy source with an antenna configuration with maximum baselines of $\sim2$ km. The different colors correspond to different regularization schemes: gray contours are for a gradient prior, and red contours are for a curvature prior \citep{warren:03}.  We see no evidence of systematic biases in the reconstructed parameters. The results shown here are typical of the mocks we have simulated. 
\label{fig:Warrens_Sim_Post}}
\end{figure*}

\section{Searching for subhalos}
\label{sec:subhalofinding_method}

Equation~\eqref{chi2} tells us how well any lens model fits the data.  That model can contain a smooth potential of the main lens, and a number of subhalos.  As we add subhalos to the mass model, the model parameter space grows considerably in dimensionality, and searching that high-dimensional parameter space can become computationally expensive.  In this section, we describe a simple approach towards expediting that search.  Our method relies on first making a linear approximation to the effect of subhalos on the model predictions, in order to identify promising points in parameter space where subhalos could significantly improve the fit.  Once we have identified those promising parameter values using the linearized search, we then conduct searches using the full nonlinear lens equations, starting at those points in parameter space.

As discussed above, lens model predictions are in general highly nonlinear functions of most of the lens parameters, and are linear only in the source parameters.
In this section, we show that once a maximum posterior model is found using the procedure described above, we can linearize the model predictions in a small local neighborhood of these fiducial parameters, even when substructure is added to the mass model.  The key aspect of low mass substructure that permits a linearized treatment, even approximately, is that the effects of substructure are perturbative when expanded in the deflection angle.

For the purpose of our discussion below, it will be useful to separate the subhalo parameters \psub\ from the rest of the parameters $\bm{p}$ in our notation.  The parameter set $\bm{p}$ therefore consists of macro-lens density parameters, background source parameters, and other nuisance parameter (e.g., antenna phases), while we reserve the subhalo parameters as \psub.  

To infer the presence of subhalos, we study the posterior PDF for the subhalo parameters, \psub,  marginalized over all the rest of the parameters $\bm{p}$, and we compare to the marginalized posterior PDFs in models with different numbers of subhalos. In particular, the ratio of the marginalized posterior of a subhalo model with mass $M_{\rm sub}>0$ to that of a model without subhalos (or equivalently, with $M_{\rm sub}=0$)  provides a route to the marginal posterior for the presence of the given subhalo. 

In our approach, we first find the maximum posterior smooth model using the procedure described in Section~\ref{sec:framework}. Next, we consider the effect of subhalos.  Note that the effects of low-mass subhalos on the lensing deflection angles entering the matrix ${\bf L}$ are quite small in general.  More precisely, low-mass substructure produces only a small perturbation to the deflection angle at every location.  This suggests a linearized treatment of subhalos \citep{dalal:02,Hezaveh13A}.
For a given subhalo (which could be parameterized by its location and mass), we add the subhalo deflections to the deflections arising from the best-fit smooth lensing potential model.
We write the marginalized posterior, $\L_m$, for the subhalo of that mass and that location as
\begin{equation}
\L_m(\psub) = \int \L (\bm{p},\psub) {\rm d}\bm{p} \propto \int e^{-\frac{1}{2} \LOGP}  {\rm d}\bm{p}
\label{marg}
\end{equation}
where \L\ is the joint posterior PDF and \LOGP\  is given by Equation~\eqref{chi2}, which we rewrite as
\begin{equation} \label{chi2linear}
\begin{split}
\LOGP(\bm{p}, \psub) =&  [\bm{D} - \bm{M}(\bm{p}, \psub)]^\T \CN^{-1} [\bm{D} - \bm{M}(\bm{p}, \psub)] \\ &  + \  \bm{p}^\T \Cp^{-1} \bm{p} \,, 
\end{split}
\end{equation}
in order to make the dependence on \psub\ explicit.
Here, \Cp\ is the prior covariance matrix of all parameters  ${\bm p}$, and \Cp\ contains the block \Cs\, as described in the previous section.
Note that as in Equation~\eqref{chi2}, our expression for \LOGP\  still includes a prior term, so that it does not solely reflect the goodness-of-fit.   The integral in Equation~\eqref{marg} is analytic when we can use a linearized treatment of parameters, i.e.,\ when subhalos are perturbative \citep{hezaveh:14a}  (also known as the Laplace approximation). To leading order, we can write 
\begin{equation}
\bm{M}(\bm{p}=\bm{p}_0+\Delta\bm{p}) = \bm{M}(\bm{p}_0) + 
\frac{\partial \bm{M}}{\partial \bm{p}}       \Delta \bm{p}
\label{MP_linear}
\end{equation}
for some fiducial parameter set $\bm{p}_0$, which we take to be the smooth model parameters from the maximum posterior model without substructure.  Substituting this expression for $\bm{M}$ into \LOGP \  gives us a quadratic expression in the parameter shift $\Delta\bm{p}$,
\begin{equation}
\LOGP(\bm{p}_0+\Delta\bm{p},\psub) = \LOGP_0 + 
2\B^\T \Delta\bm{p} + \Delta\bm{p}^\T \A^{-1} \Delta\bm{p},
\end{equation}
where
\begin{align}
\LOGP_0 &= \Delta\bm{D}^\T \CN^{-1} \Delta\bm{D}+ \bm{p}_0^\T \Cp^{-1} \bm{p}_0 \\
\A &= \left[\dOdp^\T  \CN^{-1}  \dOdp + \Cp^{-1} \right]^{-1} \\
\B &= \dOdp^\T  \CN^{-1} \Delta\bm{D}  -  \Cp^{-1} \bm{p}_0
\end{align}
and $\Delta\bm{D}=\bm{D} - \bm{M}(\bm{p}_0,\psub)$ consists of the residuals between the data and the predictions of the {\em unadjusted} smooth model plus subhalo.  Performing the integral in Equation \eqref{marg} then gives the likelihood for the subhalo parameters \psub, marginalized over all nuisance parameters $\bm{p}$,
\begin{equation}
\L_m(\psub) \propto |\A|^{-1} e^{-\frac{1}{2}\LOGP_0} \, e^{\frac{1}{2} \B^\T \A^{-1} \B} \equiv e^{-\frac{1}{2}\LOGP_m(\psub)},
\end{equation}
where we define an effective marginalized $\LOGP_m(\psub)$ in terms of the log of the marginalized likelihood,
\begin{equation}
\LOGP_m = \LOGP_0 - \B^\T\A\B + {\rm const}.
\end{equation}
Again, $e^{-0.5 \ \LOGP_0}$ is proportional to the posterior if  we assign uniform priors to the subhalo parameters and we add a subhalo to our best-fit smooth model without adjusting the parameters (i.e., for $\bm{p}=\bm{p}_0$): the $\B^\T\A\B$ term accounts for the marginalization over the smooth model parameters $\bm{p}$.
Since \A\ does not depend on subhalo parameters \psub, we can write the difference of log posterior for two models as 
\begin{equation}
\Delta\LOGP(\psub) \equiv \LOGP_m(\psub) - \LOGP_{m0}
\label{dchi2}
 \end{equation}
where $\LOGP_{m0}$ corresponds to the marginalized likelihood for the model with no subhalos.  

Because we parametrize the source emission using pixel brightnesses on a fixed grid, rather than using parameters like the locations of source clumps that can freely move, one subtlety that arises in the above expressions is that the matrix $\partial \bm{M} / \partial \bm{p}$ must involve {\em convective} derivatives rather than partial derivatives.  These convective derivatives introduce certain nonlinear terms which naively might appear to be small, but in practice we have found that neglecting the distinction between convective derivatives and partial derivatives can lead to significant errors in estimating the ability of the smooth (macro) model to compensate for the effects of substructure, in particular for the most massive subhalos.  To calculate the required convective derivatives, we evaluate the derivative $\partial \bm{M} / \partial \bm{p}$ in Equation~\eqref{MP_linear} using a deflection field that includes deflection from the subhalo model under consideration.  

Using the procedure described above, we can map out the marginalized posterior of models as a function of the subhalo parameters (e.g., location and mass).  If we find any locations within the substructure parameter space that appear to significantly improve the fit, we then initiate an optimization of the model parameters using the full nonlinear lens equations, starting from the most promising points found using the linearized search.

It is worth stressing that the linearized lens model is very accurate over most of the subhalo parameter space, as long as subhalo effects are small.  The only locations where the linearized model predictions differ noticeably from fully nonlinear model predictions are where subhalos produce significant changes to the posterior, i.e., where $|\Delta\LOGP| \gtrsim 20$, which in general only occurs within a small fraction of the possible parameter space.  Thanks to the high accuracy of the linearized approximation, we can use the maps of $\LOGP_m$ not only to identify promising starting points for our nonlinear search, but {\em also} to exclude subhalos over much of the possible parameter space where subhalos are ruled out.  In other words, we can use the maps of $\LOGP_m$ to derive constraints on the presence of subhalos, and in principle, derive constraints on the mean abundance of subhalos.  To derive bounds on the abundance, we would similarly need to map out $\LOGP_m$ for models that have 2 subhalos, 3 subhalos, and so on, for all possible subhalo locations.  However, because we are considering subhalos of low mass, their effects on the observables are perturbative and restricted to localized regions on the sky.  We will argue below that this allows us to derive approximate constraints on the subhalo abundance from maps of $\LOGP_m$ from models containing one subhalo. 

\section{implementation, simulations, and tests}
\label{sec:implement}

Perhaps the most challenging aspect of this analysis is the sheer volume of the visibility data provided by interferometers such as ALMA. For ALMA long-baseline campaign observations of SDP.81  \citep{Fomalont:15}, visibilities were typically recorded every 0.5 seconds for hundreds of baselines, each with more than 1000 spectral channels. This results in about $10^8$ visibilities for a four-hour long observation. For a reasonable pixel size, the size of the resulting Fourier operator alone can exceed a few terabytes, and fitting many models to such matrices is beyond the capabilities of medium-size clusters. Fortunately, careful binning of the visibilities can make the problem tractable. Because of the non-zero antenna size, each visibility actually samples a range of spatial frequencies around its nominal $uv$ location and visibilities within an antenna diameter of each other are highly correlated. Binning visibilities within antenna-sized patches of the $uv$-plane can reduce the size of the data dramatically (to fewer than $10^6$ visibilities per spectral channel) without significantly decreasing the information content of the data.

Even after binning, this is still a computationally challenging task: the matrices are still too large to fit in the memory of single processors, but they can fit across distributed memory on large clusters.  This requires the use of distributed linear-algebra libraries to store and manipulate these matrices. 
It is also important to note that unlike the blurring matrices used in analyzing CCD images of gravitational lenses, which are extremely sparse, the Fourier operator here is fully dense and does not permit the application of sparse libraries.
For these reasons, we take advantage of the ``\emph{Elemental}'' dense linear algebra library, which efficiently performs linear algebra operations over distributed cores \citep{Poulson:2013,PSP}.

Our pipeline includes a ``pre-processing'' part, in which the visibilities are binned according to their position in the $uv$-plane and the noise covariance matrix \CN\ is determined. The result of this pre-processing is a vector of binned visibilities which can be modeled as described in Section~\ref{sec:framework}.  

The datasets provided by ALMA contain information about the standard deviation of noise for each visibility which are proportional to the system temperature $T_{\rm sys}$. This information, however, is intended to be used for weighting of the visibilities for imaging purposes and is not adequately scaled to provide accurate root mean square (rms) noise in appropriate units (e.g., Jy).
The noise variance cannot be taken directly from the visibilities because their variation contains a significant contribution from the sky signal, which must be removed.
We achieve this by first grouping the visibilities into bins that probe the same signal, and then taking differences between visibilities that null the sky within those groups, leaving only noise. 
We can then either find the best overall scaling for the provided system temperature, $T_{\rm sys}$, to maximize the likelihood of the observed noise, or alternatively we can simply assume that the variance in the subtracted visibilities is equal to twice the noise variance. We try both methods and find that they provide consistent results.

Explicitly, if we assume that the noise rms is proportional to $T_{\rm sys}$, we can write $\CN = A^2 {\bf T}^\T {\bf T}$, where ${\bf T}$ is a diagonal matrix of system temperatures, and $A$ is an unknown scaling factor.  Assuming Gaussian noise gives
\begin{equation}
\log(\ensuremath{{\cal L}}(A)) = -\frac{1}{2} \log(|A {\bf T}|) - \frac{1}{2}  \bm{N} A{\bf T} \bm{N} + C\,,
\end{equation}
where $\ensuremath{{\cal L}}(A)$ is the likelihood of a certain value of $A$,  $\bm{N}$ is the vector of noise (the residual after subtracting half of visibilities from the other half), and $C$ is the sum of other terms that do not depend on $A$. By solving $\partial \log(\ensuremath{{\cal L}}(A)) / \partial A = 0$ for $A$ we get the most likely $A$,
\begin{equation}
A = \frac{n}{\bm{N} {\bf T} \bm{N}}\,,
\end{equation}
where $n$ is the number of visibilities used to estimate $A$ (half of all visibilities, since one half is subtracted from the other to remove the signal). We build the noise covariance matrix $\CN$ by scaling the system temperatures with this scaling. 
With appropriate variances in hand, we fit the visibilities with a smooth lens model, and search for subhalos perturbatively using the procedure described in Section~\ref{sec:subhalofinding_method}.

\subsection{Mock observations}

We have performed a large number of tests, described below, using mock data, to ensure that the pipeline provides accurate results. We have tested for errors in the noise estimation and for biases in the smooth lens model.  These simulations also quantify our false positive detection rate and subhalo detection efficiency.

\begin{figure*}
\begin{center}
\centering
\includegraphics[trim= 80 20 80 20, clip, width=0.48\textwidth]{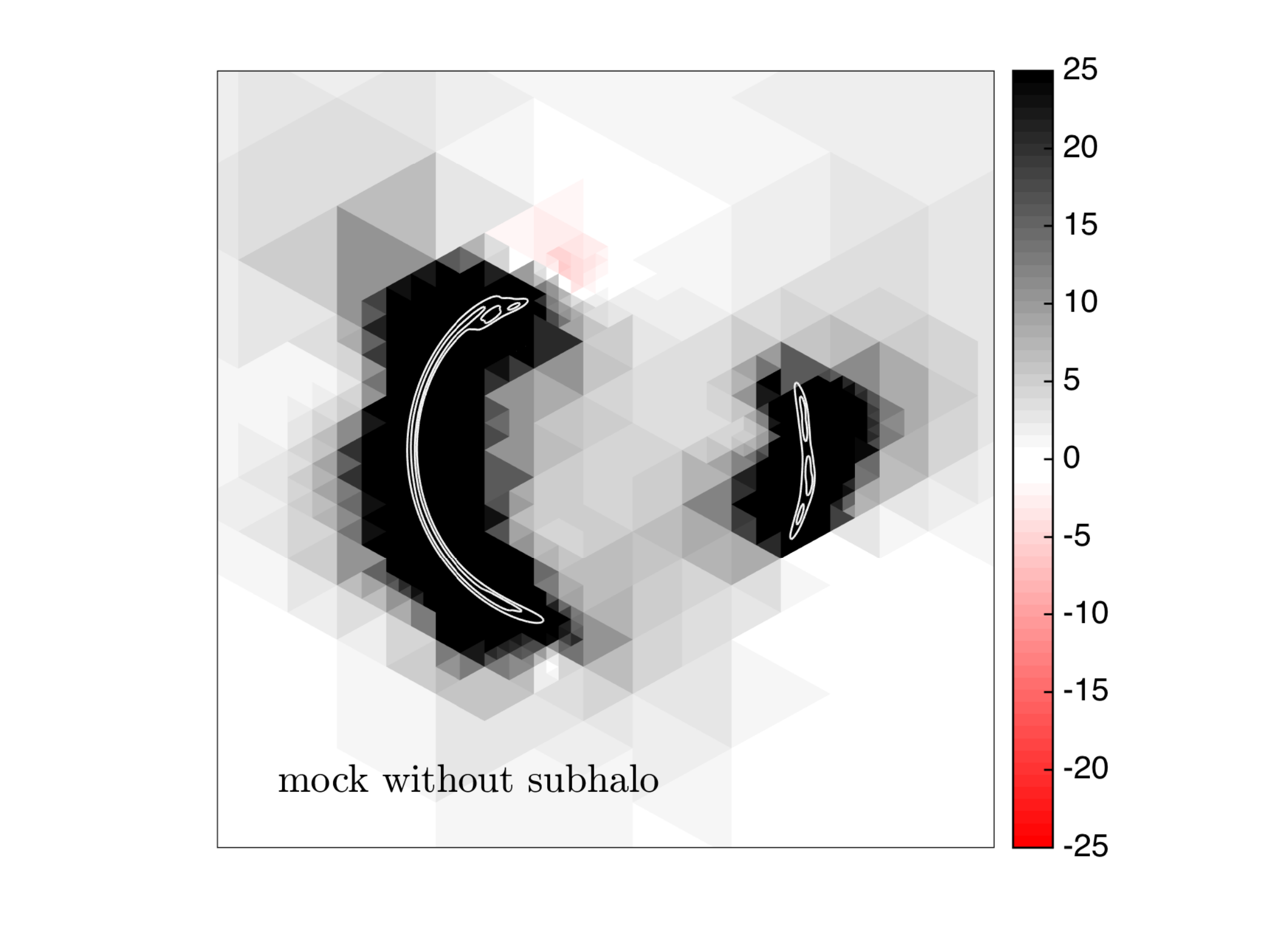}
\includegraphics[trim= 80 20 80 20, clip, width=0.48\textwidth]{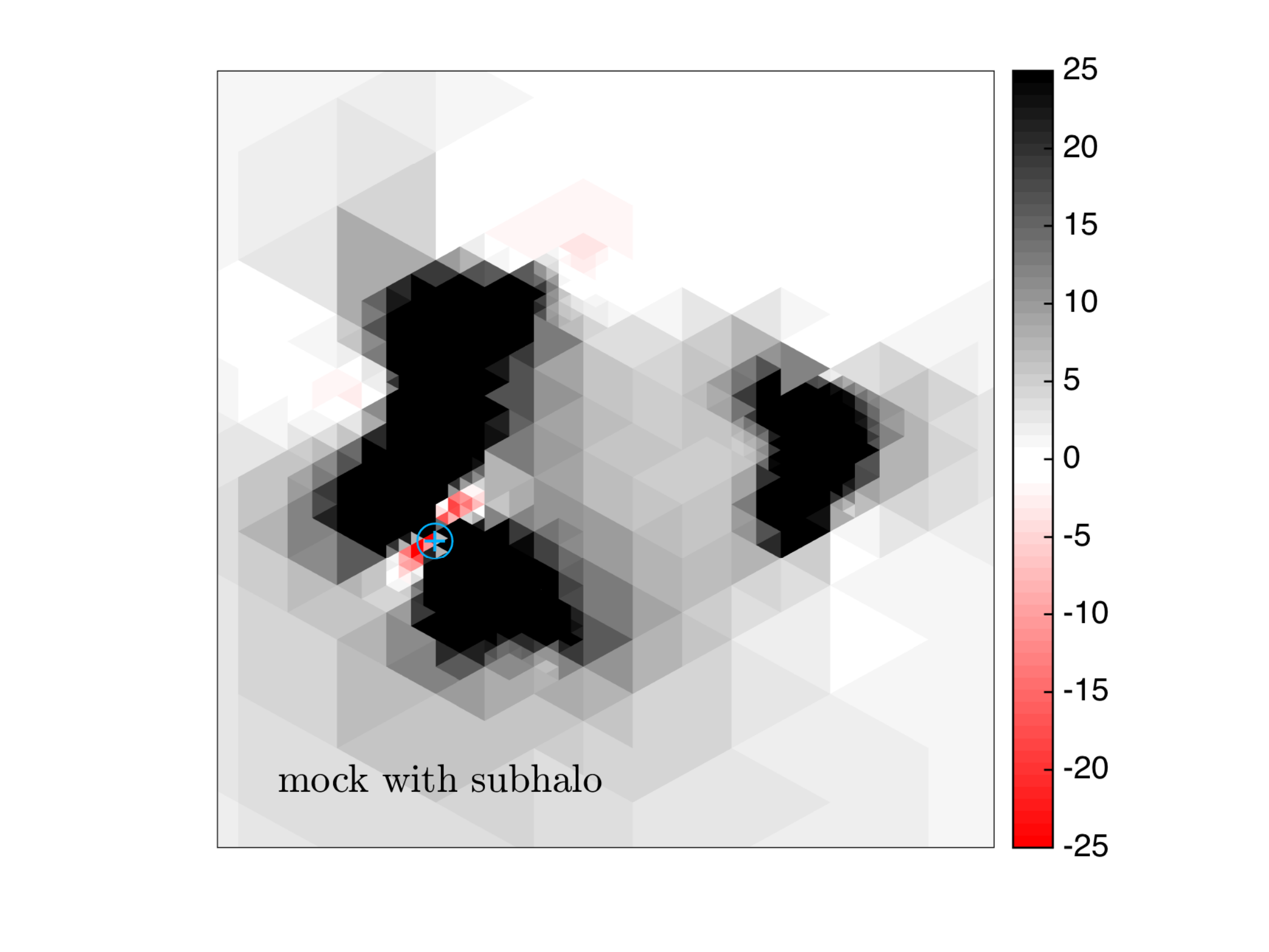}
\includegraphics[trim= 80 20 80 20, clip, width=0.48\textwidth]{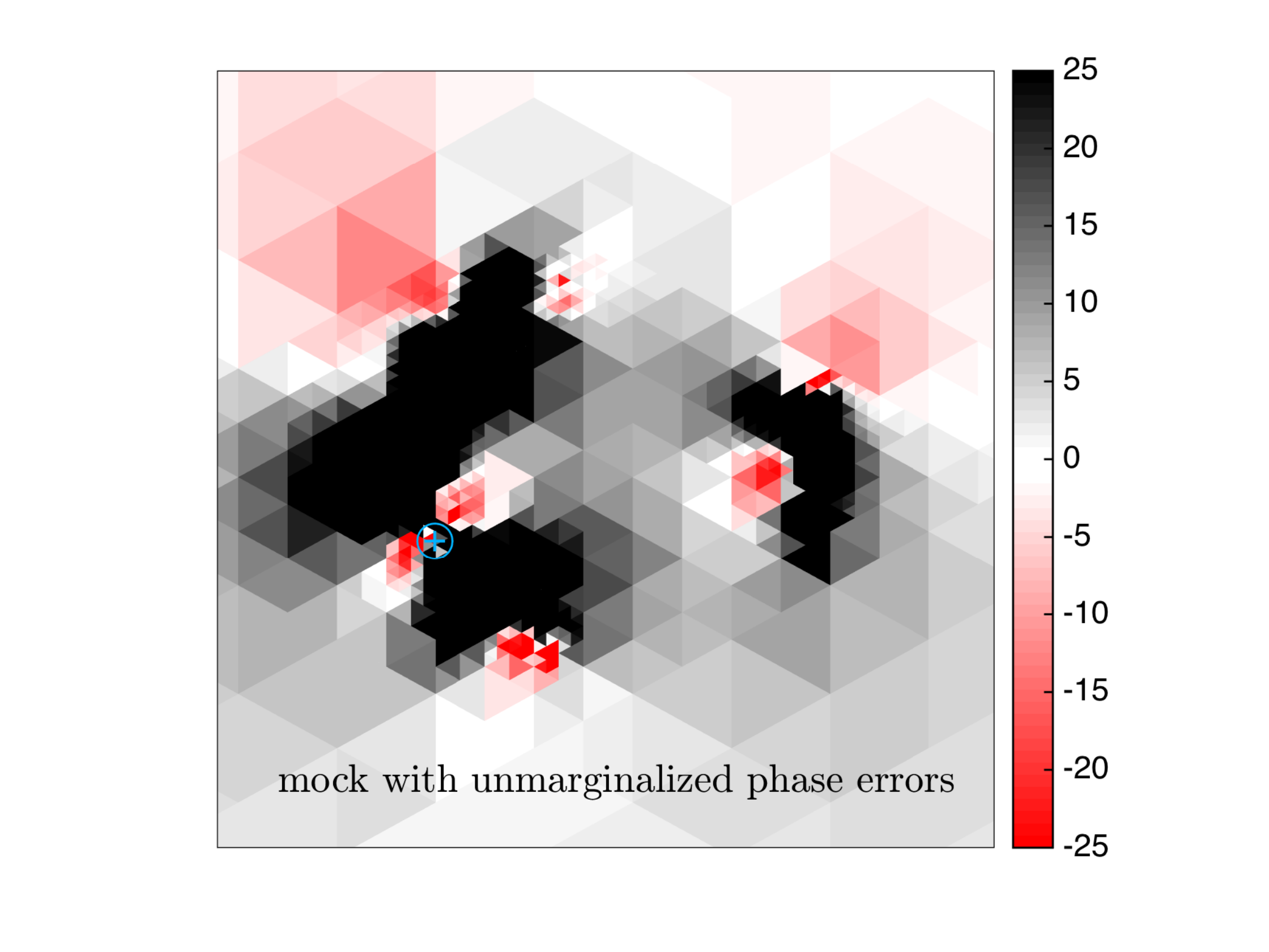}
\includegraphics[trim= 80 20 80 20, clip, width=0.48\textwidth]{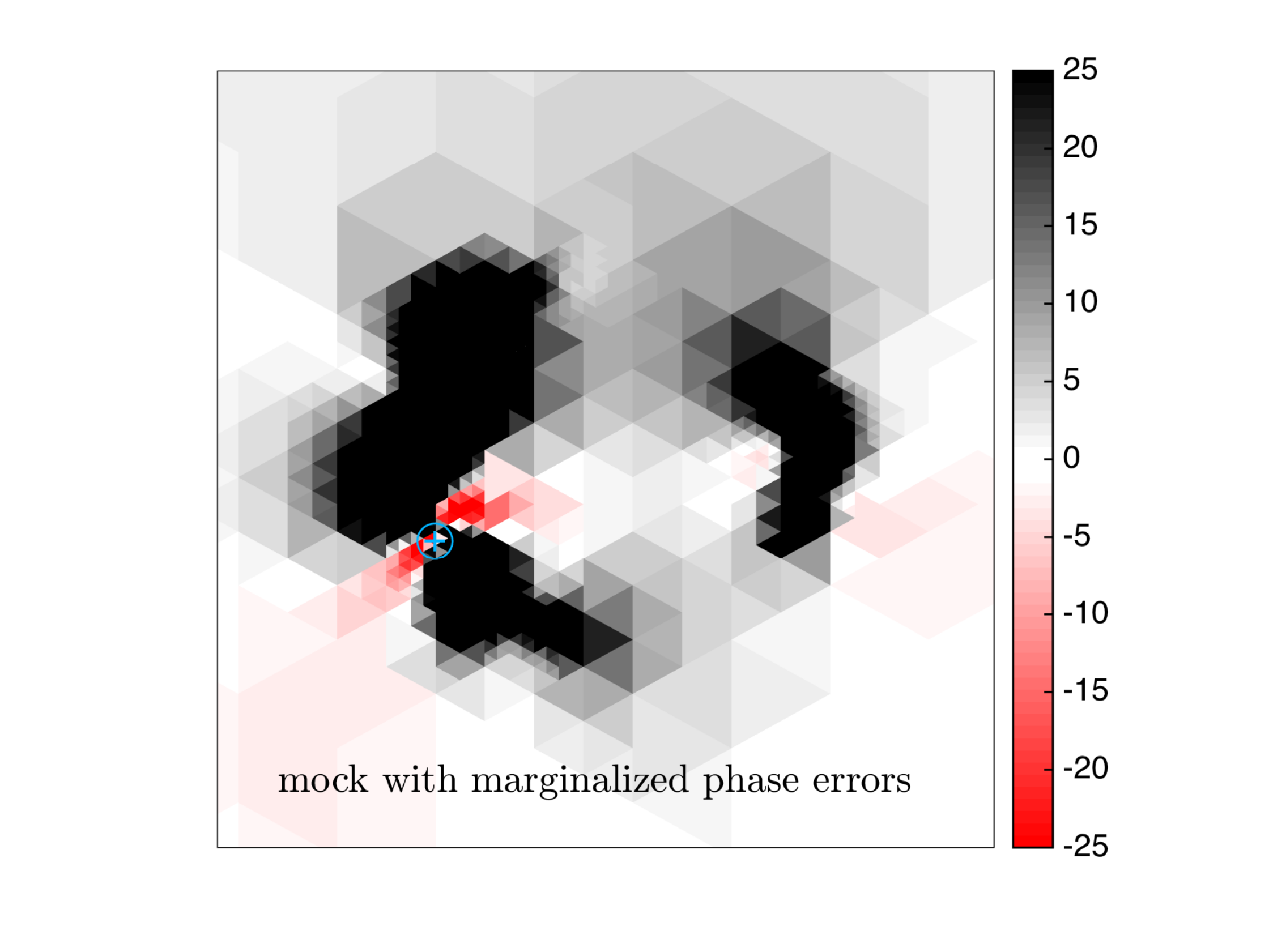}
\centering
\end{center}
\caption{ Maps of $\Delta \LOGP$ (Equation \ref{dchi2}) as a function of location, for a subhalo of mass $M=4\times 10^8 M_\odot$, using simulated data. Positive $\Delta\LOGP$ corresponds to subhalos excluded at that location, while negative $\Delta\LOGP$ corresponds to subhalos improving the log-posterior.  The mock data for the top left panel are for a simulation without any substructure, and our analysis excludes the presence of subhalos of this mass over the area of sensitivity.  The simulated arcs have been overlaid with white contours. In the remaining panels, the mocks contain a subhalo at the location of the blue circled cross. The simulation for the top right panel does not include phase errors in the measurements, the lower left panel includes such errors but does not compensate for them, and the lower right panel includes phase errors and marginalizes over them.  All four panels have the same noise realization and a model subhalo mass of $M=4\times 10^8 M_\odot$.
\label{fig:Mars}}
\end{figure*}

\begin{figure*}
\begin{center}
\centering
\includegraphics[trim= 80 20 80 20, clip, width=0.48\textwidth]{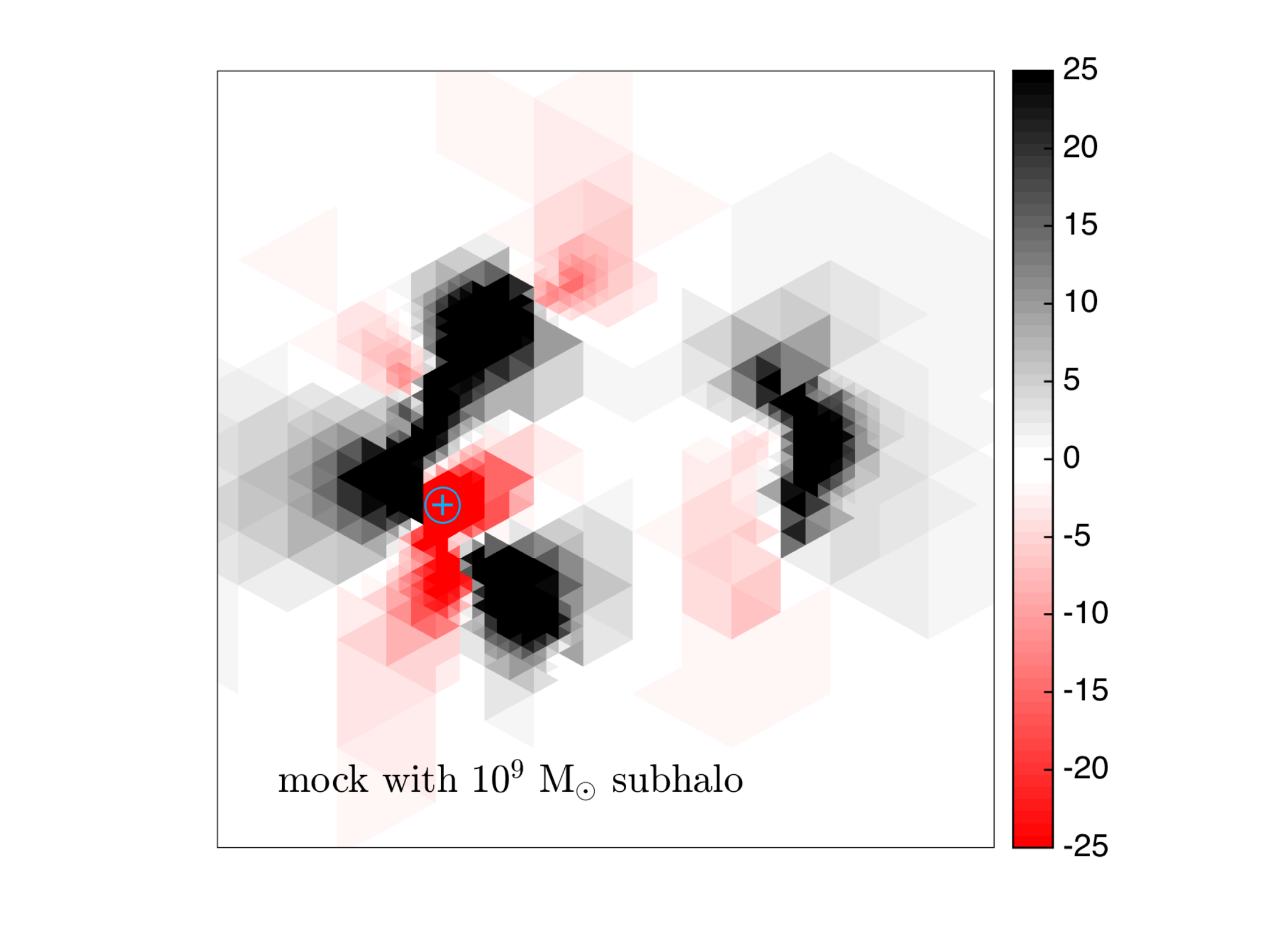}
\includegraphics[trim= 80 20 80 20, clip, width=0.48\textwidth]{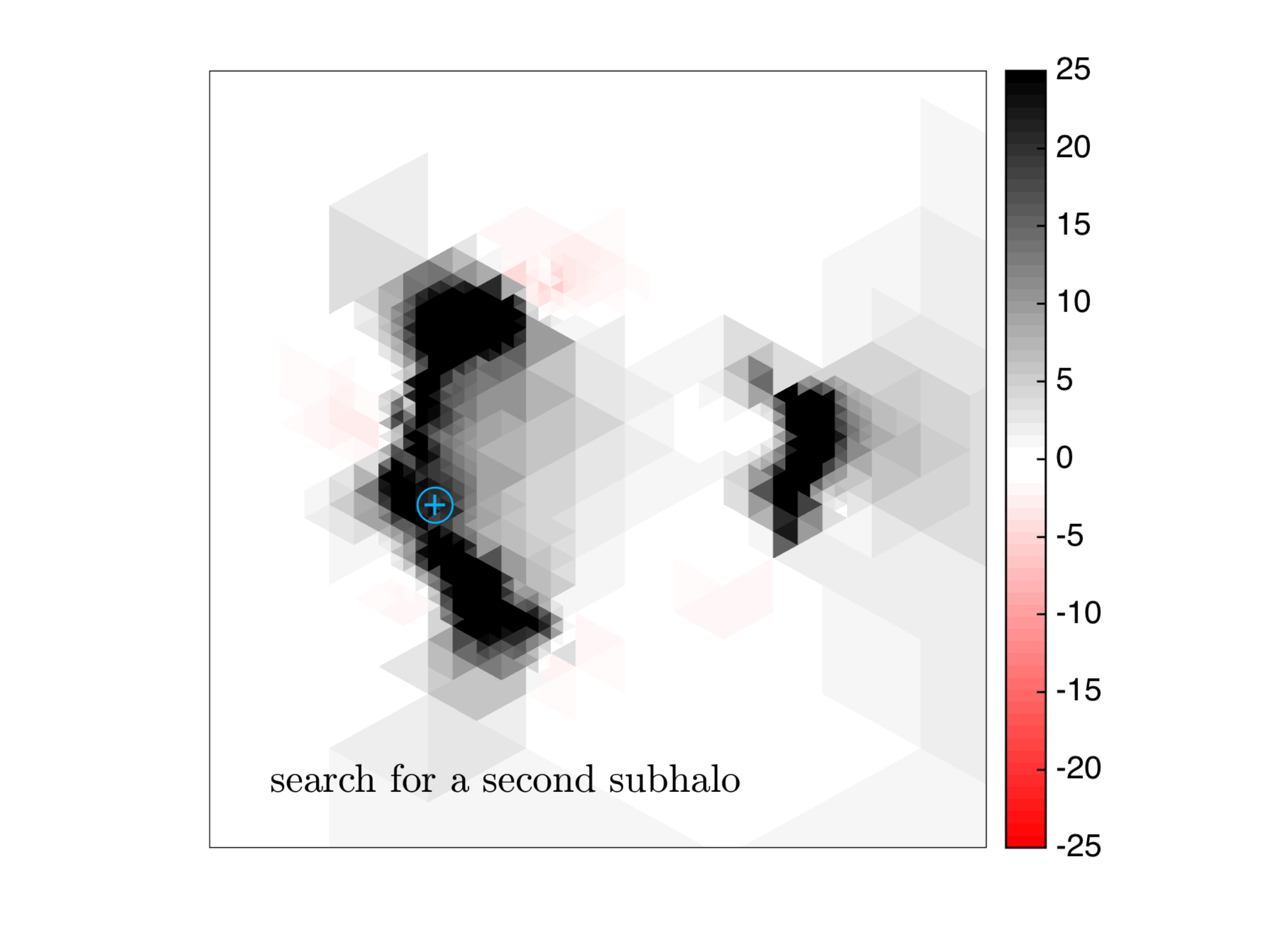}
\centering
\end{center}
\caption{$\Delta \LOGP$ maps for mock data including a massive ($M_{\rm sub} = 10^9 M_{\odot}$) subhalo, showing multiple islands where the addition of a subhalo could produce a better fit. We find that the island producing the lowest $\Delta \LOGP$ corresponds to the true position of the subhalo (blue cross). After adding the detected subhalo to the macro model and searching for a second subhalo the other islands disappear, confirming that they arise from the inability of the smooth model to fit perturbed observations at various locations on the arc. The position of the subhalo is marked with a circled cross.
\label{fig:Titan}}
\end{figure*}

The mock observations for these tests were produced using a fully independent code in order to reduce the chances of duplicating errors. For each mock dataset, a complex background source composed of tens to hundreds of smaller clumps was lensed, and the lensed images were used to simulate ALMA observations using the Common Astronomy Software Applications package (CASA). The data pre-processing steps described above (visibility binning and noise estimation) were applied to the mock data and the binned visibilities were modeled to find a best-fit smooth model.  

To explore the lens parameter space, we have implemented the Markov chain Monte Carlo (MCMC) algorithm proposed by \cite{mcmchammer}. We follow the procedure described in \cite{emcee}. Our implementation allows sampling the posterior in parallel, over many cores, taking advantage of the MPI message-passing system.  On the large computing resources that we have used \citep[e.g.,][]{XSEDE}, this allows much faster parameter explorations by evaluating multiple models simultaneously on different cores.

Figure~\ref{fig:Warrens_Sim_Post} shows the one and two dimensional marginal posterior PDFs for all lens model parameters given a mock data set. The gray shaded contours show the recovered posterior when using a gradient covariance matrix as the source prior. The red contours show the posterior for the same data, but using a curvature prior on the source parameters. The black dashed crosses show the true parameters which were used to generate the mock data. As can be seen, in both cases we successfully manage to recover the parameters with no apparent biases.

However, in the course of our tests, we encountered two issues that could result in biased parameters. The first resulted from using the power-spectrum prior as defined in Equation~\eqref{sourcePk}. Since the power spectrum does not diverge at $k\to 0$, this prior tries to minimize the total emission of the unlensed source, i.e., it biases against low magnification. This bias towards high magnification favors shallow radial density slopes. To fix this issue, we removed the DC mode from this covariance matrix, which (similar to the gradient and curvature priors) renders it insensitive to the total sum of source pixel intensities.  We can remove the zero mode of the covariance matrix by replacing
\begin{equation}
\bar{\Cs}^{-1} = \Cs^{-1} - \frac{\Cs^{-1} \, \bm{a}\, \bm{a}^\T \,\Cs^{-1}}{\bm{a}^\T\Cs \bm{a}},
\end{equation}
which follows from the Sherman-Morrison theorem.  Here $\bm{a}$ is a column vector containing only ones (i.e., the zero mode) and $\bar{\Cs}$ is the modified covariance matrix with no DC sensitivity. Removing the zero mode removes the bias towards high magnification and small source sizes.

Another potential source of bias was the choice of the area covered by  source pixels. For a given lens model, if an image pixel is mapped outside the area covered by the source grid, it will automatically be assigned a value of zero. 
A model which maps more image pixels to the defined source area may have more freedom in fitting the data compared to a model with fewer in-source-grid mappings. Therefore under equal conditions, lens models which map all image pixels to some source pixel are  preferred to models which do not. This can result in significant biases in lensing parameters (e.g., ellipticity and slope). To avoid this bias, the size of the source grid should be chosen to be large enough to cover the lensing models during the fitting process \citep{suyu:06,Vegetti:09}.

\subsection{Phase corruption}

Once a best-fit model is determined, we start the subhalo search by mapping $\Delta\LOGP$ over the subhalo parameter space (position and mass).  
As mentioned earlier, in addition to lens and source parameters, we marginalize over other nuisance parameters in the linear subhalo finder, such as antenna phase errors. This is similar to the procedure described in \cite{hezaveh:13b}, with the addition of the possibility of imposing a prior on the antenna phase errors and allowing for time-segmented phase errors.

Figure~\ref{fig:Mars} shows maps of $\Delta\LOGP$ as a function of subhalo position for various cases of simulated data. 
For a simulation with no subhalo present (top left), the analysis correctly excludes the presence of a subhalo over the region of sensitivity.
In the remaining panels, the mocks contain a subhalo at the location of the blue circled cross. The simulation for the top right panel does not include phase errors in the measurements. The subhalo is correctly detected at $\sim 5\sigma$ with no false positives. The mocks for the two bottom panels include antenna phase errors. In the bottom left panel we have not corrected for these phase errors, resulting in numerous false positives. The bottom right panel shows the results for the same mock data, when phase errors are marginalized over appropriately, resulting in the disappearance of the false detections. All four panels have the same noise realization.

This test illustrates one of the reasons why a careful analysis of the visibilities is essential to search for substructure using interferometric data.  Given the  challenges of analyzing large interferometric data sets, one might be tempted to simply analyze CLEAN images instead of visibilities.  
CLEAN images fix the value of the antenna phases and do not allow them to be marginalized over when comparing different models, which, as illustrated here, could result in spurious detections.  
In addition, CLEAN images are produced through a non-linear deconvolution procedure whose effects on the data and the correlated noise properties can not be well quantified. 
 It is worth stressing that the effects of low-mass substructure on the lensed images can be quite subtle, and so approximations and assumptions which may not lead to serious errors in other contexts could introduce significant errors for a substructure lensing analysis.  This is why we adopt the approach of explicitly modeling the visibilities and marginalizing over the nuisance parameters.

Finally, we note that although in the two right panels of Figure~\ref{fig:Mars} we only observe a single region with significant negative $\Delta \LOGP$ (where the subhalo is detected), subhalos with larger masses can result in multiple islands. Figure~\ref{fig:Titan} shows a mock generated using a simulation with a subhalo of mass $M_{\rm sub}=10^{9}M_{\odot}$. The left panel shows the $\Delta \LOGP$ map when searching for this subhalo. As can be seen, there are multiple islands  where a subhalo can produce a better fit, however the true position of the subhalo corresponds to the lowest $\Delta \LOGP$. The right panel of Figure~\ref{fig:Titan} shows the $\Delta \LOGP$ map when the detected subhalo is added to the macro parameters and a search for an additional subhalo is performed, showing that all corresponding islands disappear.

\section{Results for SDP.81}
\label{sec:results}

\begin{figure}
\begin{center}
\centering
\includegraphics[trim= 80 0 90 10, clip, width=0.5\textwidth]{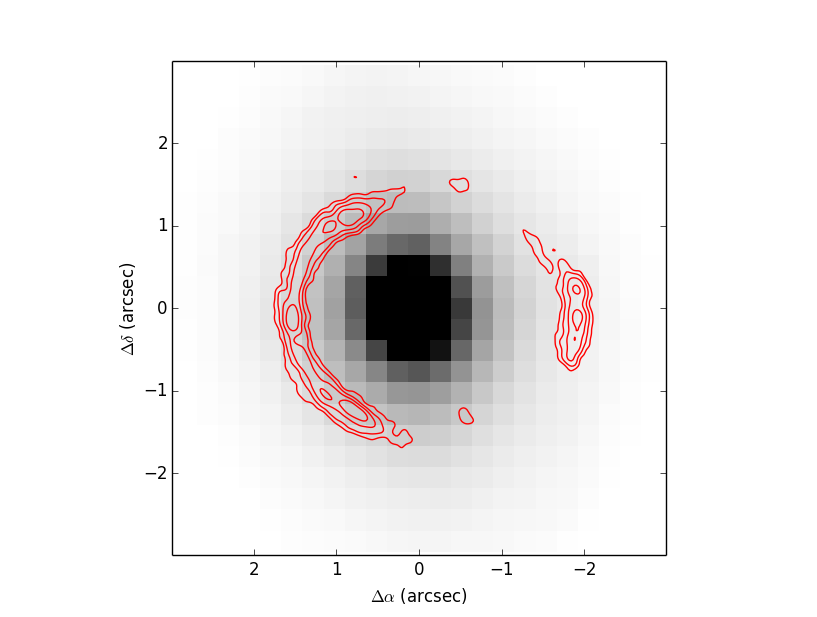}
\centering
\end{center}
\caption{The SDP.81 system.  Grayscale shows HST/WFC3 F160W data, while red contours show ALMA continuum emission in band 6. 
\label{fig:SDP81ALMAHST}}
\end{figure}

\begin{figure*}
\begin{center}
\centering
\includegraphics[trim= 125 50 85 70, clip, width=1.0\textwidth]{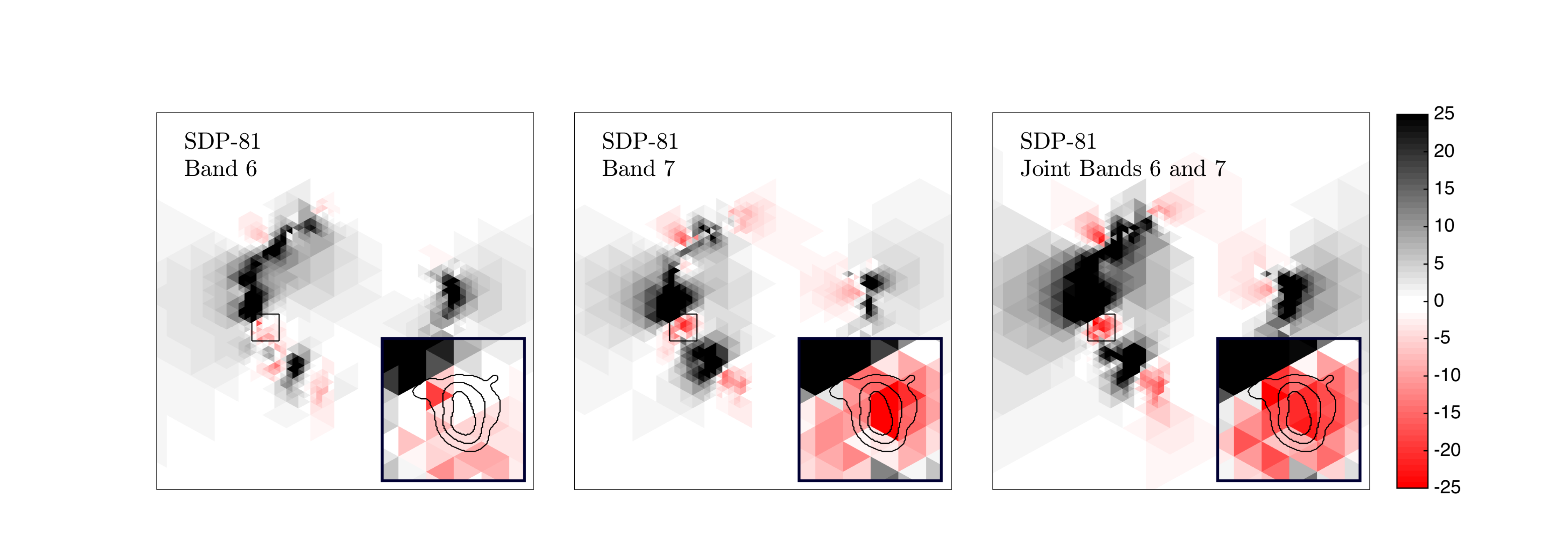}
\centering
\end{center}
\caption{Initial subhalo search using ALMA Science Verification observations of SDP.81.  Depicted are maps of linearized  $\Delta\LOGP$ from Equation~\eqref{dchi2}, showing twice the difference in log marginalized posterior probability density  between a smooth model without substructure, and a model with a subhalo of mass $M=10^{8.6} M_\odot$, as a function of location of that subhalo.  The three panels correspond to analysis of Band 6 only (left), Band 7 only (middle), and joint Bands 6 and 7 (right).  Based on the significant improvement to the fit provided by substructure (as indicated by the map), we subsequently added one subhalo to our lens model, and re-optimized the model parameters (see Table~\ref{lens_pars}). The contours in the insets show the 1-, 2-, and $3-\sigma$ confidence regions for the position of the subhalo from a non-linear joint fit to the data.
\label{initial_search}}
\end{figure*}

\begin{figure*}[ht]
\begin{center}
\centering
\includegraphics[trim= 110 50 100 50, clip, width=0.90\textwidth]{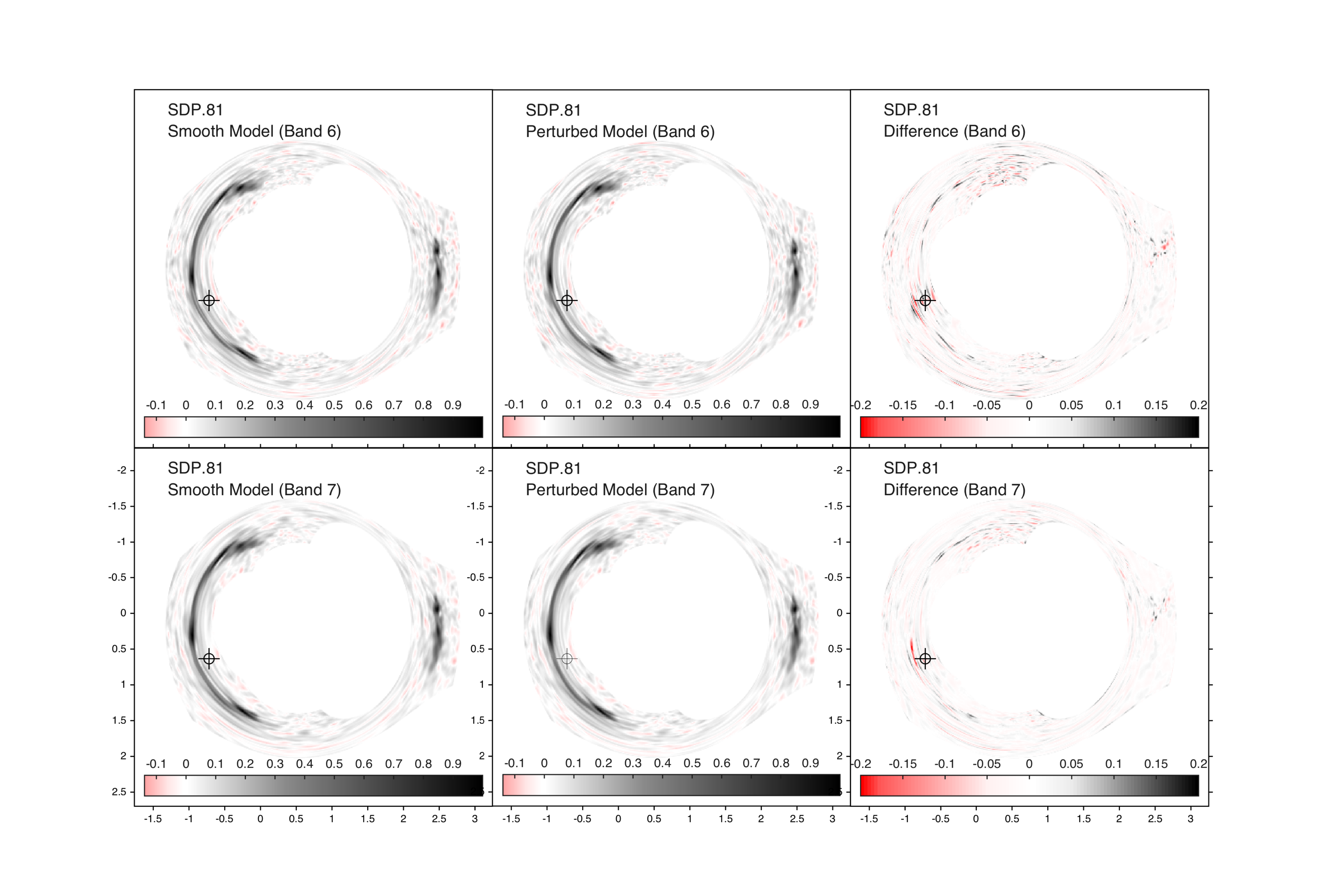}
\centering
\end{center}
\caption{ The top left panel shows the sky emission model in band 6 for the best-fit smooth lens parameters for the SDP.81 data. The top middle panel shows the same for the perturbed model and the top right panel the difference between the two models. The bottom panels show the same for band 7. The bright feature in the difference plots is mainly caused by the astrometric anomaly of the arc.
\label{fig:model_res}}
\end{figure*}

\begin{figure*}[t]
\begin{center}
\centering
\includegraphics[trim= 50 80 70 70, clip, width=1.0\textwidth]{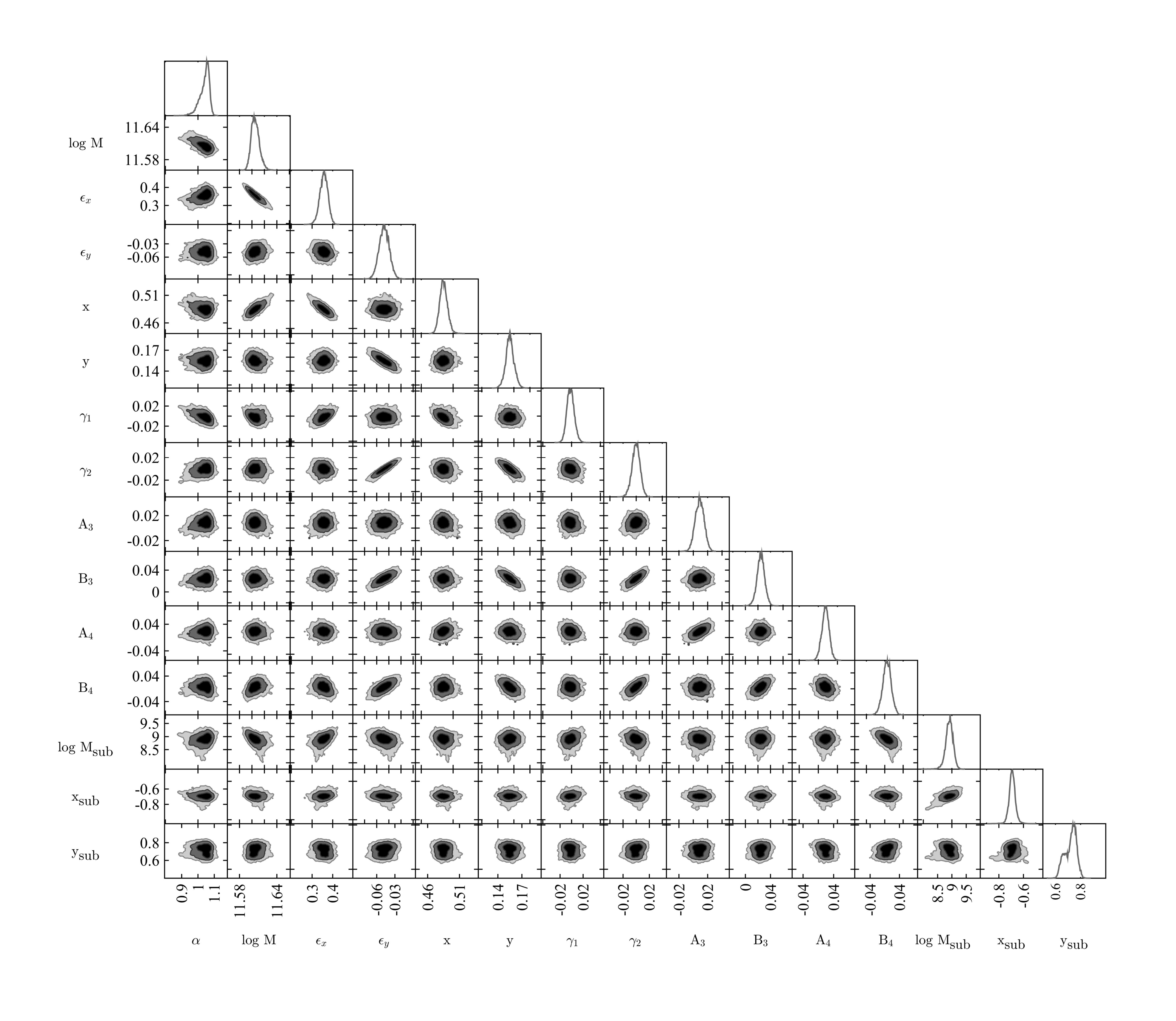}
\centering
\end{center}
\caption{ Contours show the 1-, 2-, and $3-\sigma$ (68.3, 95.5, and 99.7\%) confidence regions for the parameters in our lens model for a joint analysis of bands 6 and 7 data. 
\label{MCMC_errors}}
\end{figure*}

\begin{figure}
\begin{center}
\centering
\includegraphics[trim= 20 20 10 70, clip, width=0.5\textwidth]{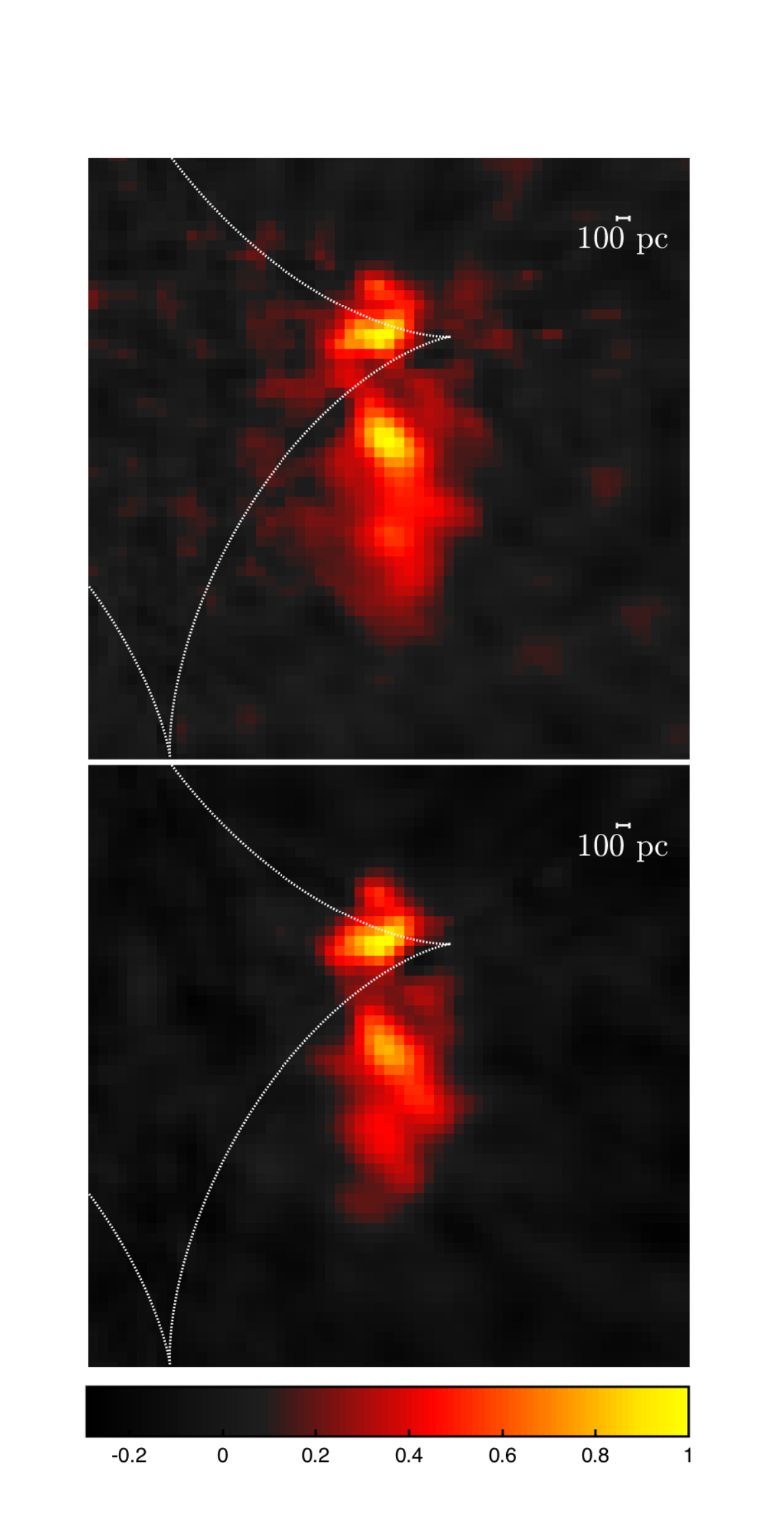}
\centering
\end{center}
\caption{Reconstructed source continuum emission from Band 6 (top panel) and Band 7 (bottom panel) data on a 10 milli-arcsec pixel grid.  The white dashed curve shows the tangential caustic predicted by our best-fit smooth model. 
\label{fig:rec_src}}
\end{figure}

\begin{figure}
\begin{center}
\centering
\includegraphics[trim= 20 20 20 20, clip, width=0.48\textwidth]{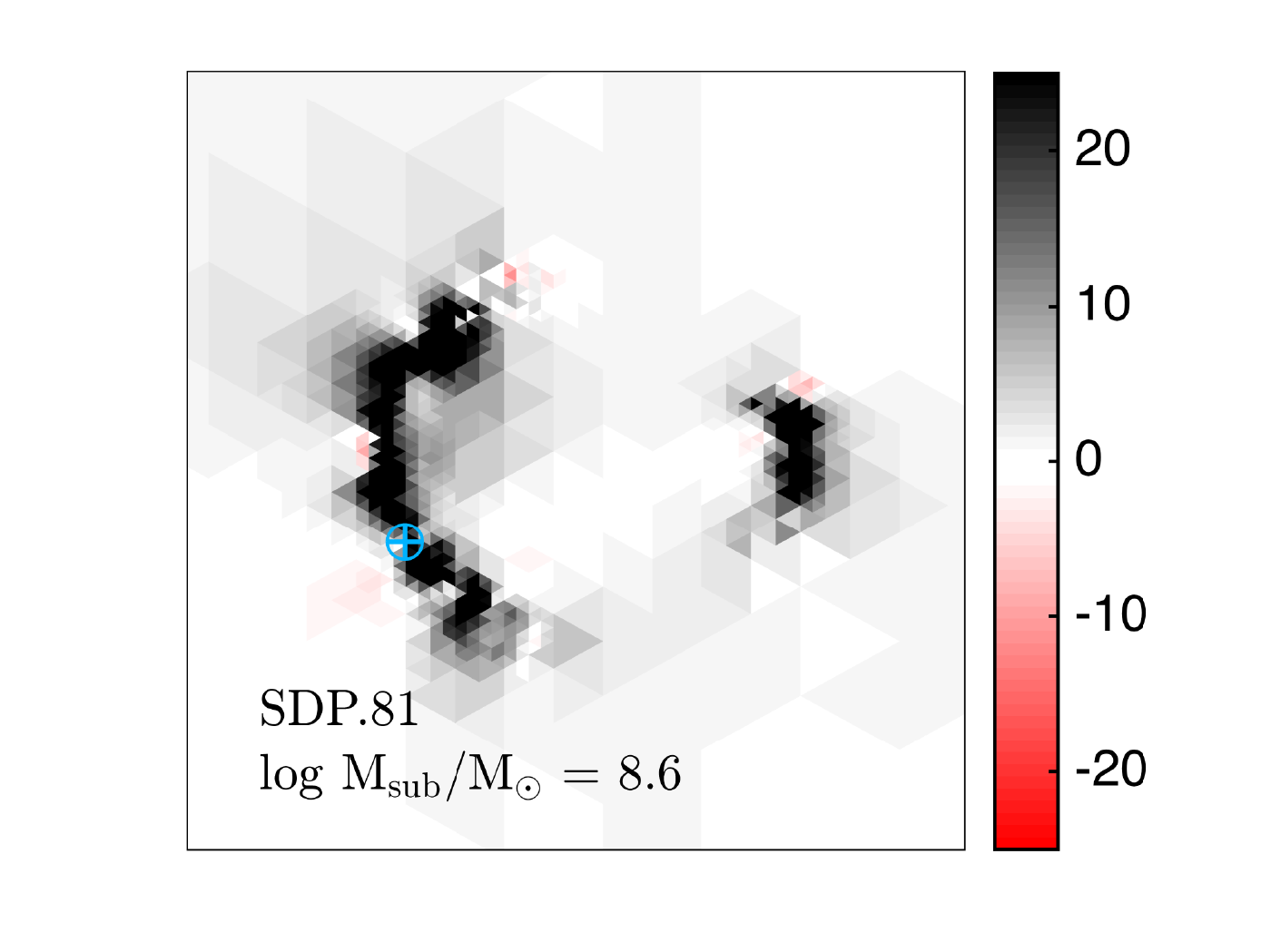}
\includegraphics[trim= 20 20 20 20, clip, width=0.48\textwidth]{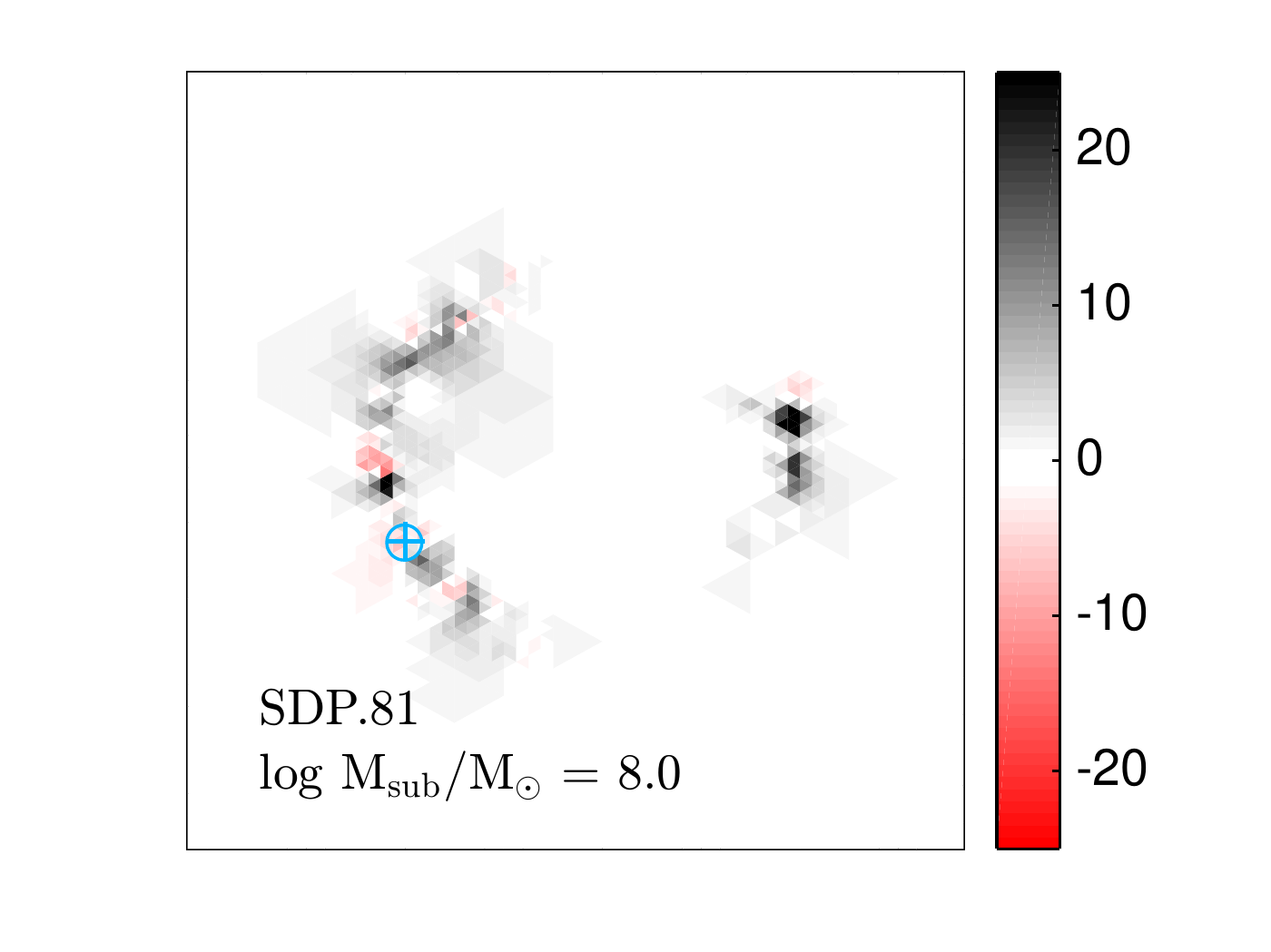}
\centering
\end{center}
\caption{Search for additional substructure.  Top panel shows a map of linearized $\Delta\LOGP$ for a second subhalo, of mass $M=10^{8.6} M_\odot$, following the inclusion of one subhalo of mass $M\approx10^9 M_{\odot}$ at the location of the blue circled cross, using a joint analysis of bands 6 and 7.  After adding the subhalo to the main lens model, no additional subhalos of this mass are found. The bottom panel shows a similar analysis for a lower mass subhalo, showing a marginal improvement of  $\LOGP$ at another point near the first detection.
\label{secondpass}}
\end{figure}

We use the recent ALMA Science Verification observations of the strongly lensed system SDP.81 \citep{SDP81} to illustrate the application of this method to real data. Our analysis shows that this dataset is highly sensitive to the effects of low mass subhalos.  In the analysis presented here, We used only the calibrated continuum data from bands 6 and 7. In future work we will present our analysis of the full dataset, including CO line data.  Figure \ref{fig:SDP81ALMAHST} shows the lensing galaxy observed by the \emph{Hubble Space Telescope} (greyscale) superimposed on the lensed arcs observed by ALMA.

We estimate the noise variance for each visibility using the procedure described above and bin the visibilities in 12 meter cells. However, we only bin visibilities that share the same baseline and are taken within a short ($\lesssim 20$ min) observing period. This allows us to assign a separate antenna phase error to each baseline at different time intervals (since the phase errors could slowly change during the course of an observation). This results in $\sim0.5$ million binned visibilities, a factor of $\sim 20$ fewer than the original $\sim10$ million. We note that the shorter ($\lesssim 2$ km) baselines in the binned visibilities have significantly higher signal-to-noise compared to the longer baselines because (a) for resolved sources, the signal diminishes on the longest baselines, and (b) the visibilities are sampled more densely on short baselines, meaning that the short-baseline bins contain a larger number of  visibilities. This allows us to speed up our search for the best-fitting parameters using the following approach.  We first use only the subset of baselines shorter than 2 km in our initial MCMC analysis to localize the neighborhood of the best fit.  
This approach greatly expedites our MCMC optimization, since many evaluations of the likelihood are required to fully search the highly multidimensional parameter space of our smooth model.
Once this initial localization has been achieved, we can use the full dataset. We note, however, that the observations of SDP.81 are performed on various dates which span many weeks, with significantly variable observational conditions, giving rise to a large range in noise properties. We, therefore, decide to model binned visibilities with noise lower than 5.0 and 5.5 mJy in bands 6 and 7 respectively, which contain about 40\% of the total number of unbinned visibilities. 

Our smooth density model consists of the following terms.  First, the main lens is described by a singular elliptical power law surface density profile of the form $\kappa(x,y) \propto [x^2+y^2/(1-\epsilon)^2]^{-\alpha/2}$ where $\alpha$ is the radial power-law index, and $x$ and $y$ are measured relative to the lens centroid $x_{\rm lens}, y_{\rm lens}$ \citep{Barkana:98}. To avoid degeneracy of the orientation angle when ellipticity is close to zero, we use fitting parameters $\epsilon_x$ and $\epsilon_y$, defined so that $\epsilon = (\epsilon_x^2+\epsilon_y^2)^{1/2}$ and the orientation angle is given by the arctangent of these components. Additionally, we allow for low order angular multipoles in the main lens, of the form $\kappa_m(r,\phi) =[A_{m} \cos(m\phi) + B_m \sin(m\phi)](r/r_s)^{-\alpha}$ for $m=3,4$, where $r_s = 1^{\prime\prime}$. Note that the same radial slope $\alpha$ and centroid $(x_{\rm lens}, y_{\rm lens})$ are used for the multipoles and for the ellipsoidal piece. Finally, we also allow for external shear, parameterized by the usual components $\gamma_1$ and $\gamma_2$.  Overall, therefore, our primary lens model contains 12 freely adjustable parameters.

\subsection{Initial subhalo search}

Once a smooth model is obtained, we use the best-fit parameters to perform a linearized search for subhalos. As we have mentioned above, these lens parameters, source parameters, antenna parameters, etc., all become nuisance parameters that we marginalize over for every different model when we search for subhalos.  We follow \citet{Hezaveh13A} and model the subhalo deflection field using a truncated isothermal surface density profile, also called a pseudo-Jaffe profile \citep{Munoz01}.  This profile is characterized by a velocity dispersion $\sigma_v$ and truncation radius $r_t$, and the total mass of the subhalo is given by $M_{\rm sub} = \pi \sigma_v^2 r_t/G$.  To reduce the dimensionality of the subhalo parameter space, we assume that $r_t$ is related to $\sigma_v$ by $r_t = (\sigma_v/\sqrt{2}\sigma_G)r_{\rm E}$, where $\sigma_G$ is the velocity dispersion of the main lens, determined from its observed Einstein radius $r_{\rm E}$.  We search for subhalos over a range of subhalo masses, over a $8\times8$ arcsec area around the lens center. Figure~\ref{initial_search} shows the results of our initial search.  The figure plots $\Delta\LOGP$, twice the difference in marginalized log posterior between a model with a subhalo compared to our smooth model, as a function of subhalo location for a subhalo mass $M=10^{8.6} M_\odot$.  As the figure indicates, there are several locations where adding a subhalo improves the posterior considerably, with the most significant having $\Delta\LOGP=-22.2$. 

As discussed above in Section~\ref{sec:implement}, improper modeling of systematics and unknown errors can lead to spurious detections of substructure.  We have attempted to mitigate these effects by marginalizing over many potential systematics, including time-varying antenna phase errors.  Nevertheless, it is possible that the apparent detection of substructure indicated in Figure~\ref{initial_search} could be due to an unknown interferometric data corruption, such as visibility decorrelation or rapidly varying antenna phase errors.  
Given that such errors are temporally variable, an analysis of multiple datasets observed at different times can reveal if our analysis is affected by them.
As a test of this, we analyzed bands 6 and 7 data separately, noting that they were obtained on different dates.
Our analysis reveals a consistent pattern between the two bands (see Figure~\ref{initial_search}), giving us confidence that the level of unknown systematics from such effects is below our statistical uncertainties. Figure~\ref{fig:model_res} illustrates the difference between our best-fitting model without substructure and the best-fitting model with substructure for bands 6 and 7. As expected, the subhalo's effect is largely localized to its immediate vicinity and the counter-images of that region.

Based on the results of this initial linearized search, we then expanded our lens model to include a subhalo, with 3 adjustable parameters: mass $M_{\rm sub}$, and 2-D location $\bm{x}_{\rm sub}$.  We then re-fit the joint data set, re-optimizing all the parameters fully nonlinearly.  We find that a model with a subhalo of mass $M=10^{8.96} M_\odot$ improves the marginalized log posterior fit by $\Delta\LOGP = -47.3$ in the joint fit (note that the initial linear search was performed at $ M_{\rm sub}=10^{8.6} M_{\odot}$).  
Based upon this result, we conclude that the ALMA Science Verification observations of SDP.81 detect a subhalo in the projected mass distribution. Having found the best-fit parameters for the detected subhalo, we then sample the full parameter space  (smooth lens and subhalo parameters) non-linearly using our Markov Chain Monte Carlo sampler. Figure~\ref{MCMC_errors} shows the error covariance of the reconstructed lens parameters for the joint fit to bands 6 and 7.
We do not find evidence for significant degeneracies between the subhalo parameters and the parameters of the smooth lens model, including low-order multipoles in the gravitational potential.  This confirms  findings that such multipoles cannot mimic the effects of small-scale substructure for lenses with high-quality arcs \citep{KochDalal04}.  

The full set of best-fit lens model parameters are presented in Table~\ref{lens_pars}. Many previous works have modeled the lens potential in SDP.81, using HST data \citep{Dye:14}, Submillimeter Array data \citep{Bussmann:13}, and ALMA data \citep{Dye2015,rybak:2015a,wong2015,tamura2015,hatsukade2015,rybak2015b}. 
Our smooth model has a larger ellipticity compared to these models. We note however that our model has more degrees of freedom (e.g., angular multipoles) and phase errors, and that the degeneracy of some of these additional parameters with ellipticity may shift its value. We do find that models with parameters given by these authors produce reasonable fits to the data. We also performed the linear subhalo search for these parameters, finding that they produce similar results and that the conclusion of the presence of the subhalo is robust against these variations. Figure~\ref{fig:rec_src} shows the reconstructed source using this model with pixel size of 10 milli-arcsec in band 6 (top panel) and band 7 (bottom panel). 

This model appears to be a good fit to the data, when we fit the entire data set.  The full data set, however, includes emission unrelated to SDP.81.  The ALMA primary beam covers approximately $\sim 25^{\prime\prime}$, of which only the central few arcseconds are relevant for strong lens modeling.  If we model the sky emission only over a $5\times 5$ arccsec area centered on the lens,  our model obtains $\chi^2 = 2\times 10^5$ for $1.7\times 10^5$ degrees of freedom, suggesting that not all the signal in the data has been modeled. However, if we expand our source-plane image to cover the entire primary beam, additional flux is indicated away from the lensed galaxy and the $\chi^2$ decreases to $1.7\times 10^5$. 
Since this emission originates from regions well separated from the lensed images (far beyond the correlation length of the dirty beam), it has no model covariance with the lens parameters, and we therefore neglect it in the remainder of our analysis.

\begin{table}[tp]
\begin{centering}
\caption{Best-fit lens parameters with 68\% uncertainties}
\label{lens_pars}
\begin{tabular}{lll}
{\bf Parameter}    &  {\bf Definition}     & {\bf Value}       \\ \hline
$\alpha$            &  radial slope      & $1.06  \pm 0.03$    \\
$\log_{10} (M_{\rm 10 kpc})$ & mass within 10 kpc ($M_\odot$)    & $11.60 \pm 0.006$   \\
$\epsilon_x$              & ellipticity  x  & $0.371 \pm 0.019$     \\
$\epsilon_y$              & ellipticity  y & $-0.046 \pm 0.008$     \\
$x_{\rm lens}$              & lens x ($^{\prime\prime}$)  & $0.481 \pm 0.006$     \\
$y_{\rm lens}$              & lens y ($^{\prime\prime}$)  & $0.154 \pm 0.005$     \\
$\gamma_1$                 & external shear   & $0.0004 \pm 0.006$ \\
$\gamma_2$                 & external shear   & $0.0017 \pm 0.006$ \\
$A_{3}$          & m=3 multipole   & $[5.90 \pm 6.26]\times 10^{-3}$ \\
$B_{3}$          & m=3 multipole   & $[25.44 \pm 6.00]\times 10^{-3}$ \\
$A_{4}$          & m=4 multipole   & $[12.53 \pm 10.10]\times 10^{-3}$ \\
$B_{4}$          & m=4 multipole   & $[6.52 \pm 11.20]\times 10^{-3}$ \\
$\log_{10} (M_{\rm sub})$          & subhalo mass ($M_\odot$) & $8.96 \pm 0.12$ \\
$x_{\rm sub}$          & subhalo position x ($^{\prime\prime}$) & $-0.694 \pm 0.025$ \\
$y_{\rm sub}$          & subhalo position y ($^{\prime\prime}$) & $0.749 \pm 0.044$ 
\end{tabular}\par
\end{centering}
\medskip
\textsc{Table}~\ref{lens_pars}.--- Table of best-fit parameter values from a joint fit to bands 6 and 7 data.  Positions are in arcseconds relative to the ALMA phase center.
\end{table}

\subsection{Search for additional substructure}

\begin{figure*}[ht]
\begin{center}
\centering
\includegraphics[trim= 30 40 0 0, clip, width=1.00\textwidth]{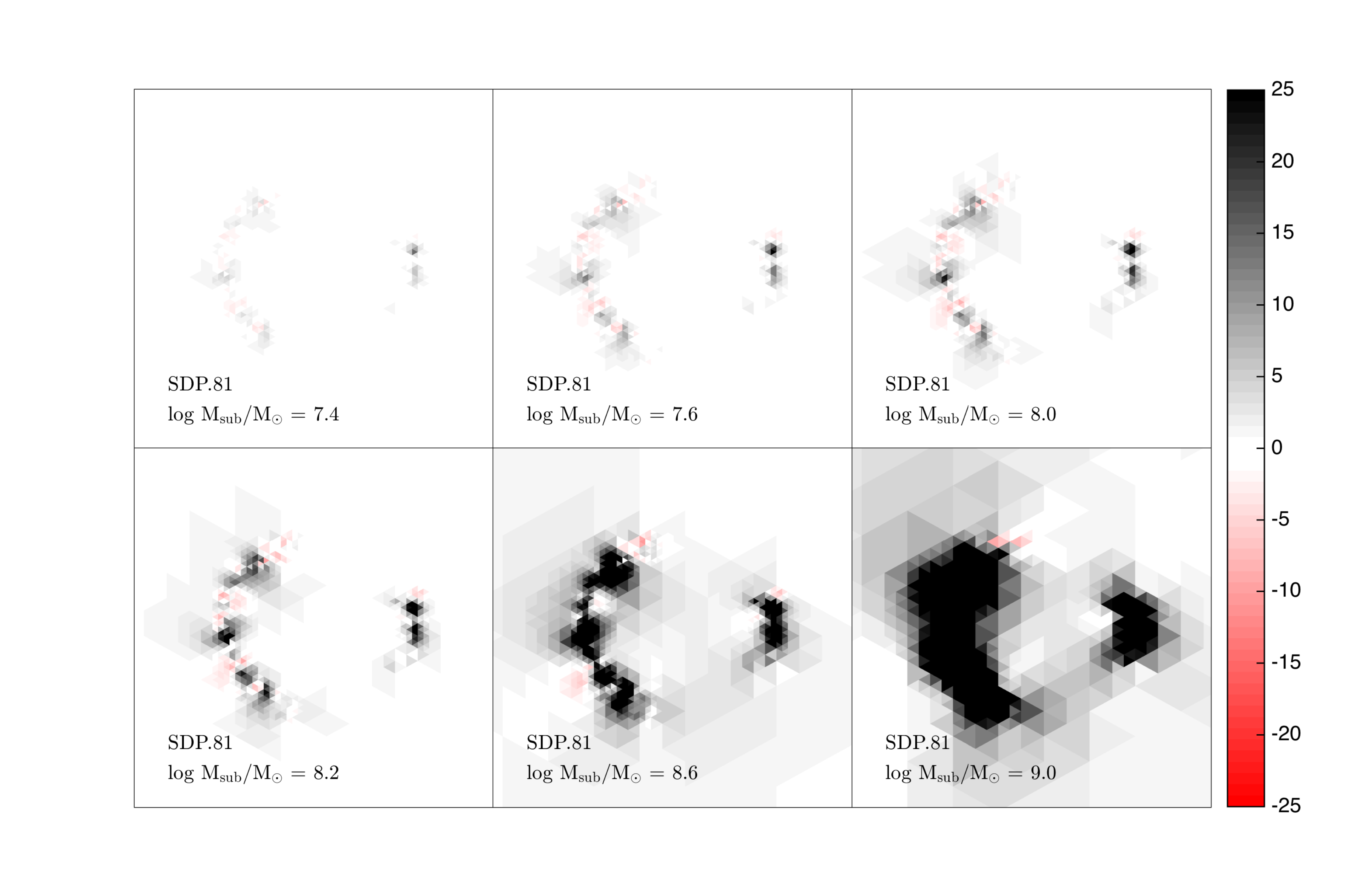}
\centering
\end{center}
\caption{ Maps of $\Delta\LOGP$ for a 3rd subhalo. When two subhalos are included in our main lens model, no significant evidence for any additional substructure is found in the joint band 6 and 7 analysis.  
\label{fig:subhalo_chi2_maps}}
\end{figure*}

ALMA observations of SDP.81 allow us to search for additional substructure besides the subhalo detected in the previous subsection.  Given our lens model (including one subhalo of $M_{\rm sub}=10^{8.96} M_{\odot}$), we next searched for additional substructure using the linearized treatment discussed in Section~\ref{sec:subhalofinding_method}.  We repeated our search for a second subhalo, by linearly expanding about a smooth model now containing a subhalo of mass $M_{\rm sub}=10^{8.96} M_{\odot}$.  As before, we marginalize over all parameters of the smooth model, including the mass and location of the detected subhalo discussed above.

The inclusion of the subhalo in our main lens model removes any improvement to the marginalized posterior from additional subhalos of mass $M_{\rm sub}\ge 10^{8.6} M_\odot$, as illustrated in top panel of Figure~\ref{secondpass}.  Instead, additional subhalos of this mass are excluded from occurring near the observed arcs.  For lower subhalo masses, however, we find that there are certain locations where addition of a second subhalo can improve the marginalized posterior.  The lower panel of Figure~\ref{secondpass} illustrates this for $M=10^{8} M_\odot$.  The red regions in the map depict the locations where a subhalo of this mass can improve the marginalized posterior.  Based on the improvement suggested in this figure, we attempted adding a second subhalo to our main lens model.  Our nonlinear fit found that a subhalo of mass $M\approx 10^8 M_\odot$ could improve the fit marginally, with $\Delta\LOGP \approx -22$, compared to the 1-subhalo model.  

We are hesitant to consider this a detection of a second object, however.  First, the detection is marginal $(<5\sigma)$.  Also, the second object is spatially near the $M_{\rm sub}=10^{8.96} M_\odot$ subhalo.  This suggests that the second object may possibly be an artifact of our modeling.  We have assumed an extremely simple mass profile for the subhalos, a spherical tidally truncated pseudo-Jaffe model, which is unlikely to describe realistic subhalos in great detail.  Small deviations in the actual mass profile compared to our model could show up as residuals that could be fitted by lower mass substructure.  On the other hand, this object could be real.  The close proximity to the first subhalo might arise simply because our area of sensitivity to detecting subhalos is relatively narrow.  Given these ambiguities, we do not label the second object as a detection, however we do include it in our main lens model.  Inclusion of additional data, in particular the CO lines which we have not used in this analysis, may be able to determine conclusively whether a lower mass subhalo is present.  To avoid biases arising from assuming a specific subhalo profile, it may be advantageous to reconstruct  the substructure density field with more flexible models.  For example,  a pixelated substructure map \citep[e.g.,][]{Vegetti:09} would allow more freedom in the inferred substructure density field.  Given the quality of the ALMA SDP.81 data set,  such an approach may be warranted.

Following the inclusion of this second subhalo in our lens model, no significant evidence for additional substructure is found in the dataset (see figure\ \ref{fig:subhalo_chi2_maps}).

\subsection{Bounds on the subhalo mass function}

Our modeling of the mass distribution around SDP.81 tells us where subhalos appear to be present, as well as where subhalos appear to be excluded, and by combining those two constraints we can derive bounds on the mean abundance $n(M)$ of dark matter subhalos in the vicinity of SDP.81.  Specifically, we use our nonlinear mass model to tell us the number and masses of the detected subhalos, as well as the area over which those subhalos are found, and we use our linearized $\Delta \LOGP$  \ maps to determine the area on the sky where subhalos are excluded.  Each piece (detections and exclusions) gives a likelihood for the mean number density of subhalos, $P[n(M)]$, and by multiplying the likelihoods derived from the detections and from exclusions, we derive our full constraints on the mass function.

Calculation of the constraints from significantly detected subhalos is a straightforward application of Poisson statistics.  The constraints from non-detections are slightly more complicated, due to our spatially varying sensitivity.  If we had a well-defined area over which subhalos were definitively excluded, then computing the Poisson statistics from that area would be straightforward.  Instead, our sensitivity to substructure varies considerably over the area of interest.  There are many pixels in our $\Delta \LOGP$ maps that slightly disfavor the presence of subhalos, and collectively those weak constraints over many pixels can combine to permit interesting bounds on the abundance.  Below, we describe the method we use to determine upper limits on the subhalo abundance from our linearized $\Delta \LOGP$ maps.

We assume that the incidence of subhalos is a Poisson process with mean projected number density $n(M)$ that is spatially uniform across the region over which we are sensitive to the effects of subhalos.
In principle, this allows us to use Bayes' theorem to derive the likelihood of abundance $n(M)$ given our observed data, $P(n|{\rm data}) \propto P({\rm data}|n)$.  To estimate $P({\rm data}|n)$, we can sum over all possible realizations of subhalo positions and masses, weighted by the Poisson probabilities for realizing each possible configuration.  For clarity, we will first derive the likelihood of density $n(M)$ for indistinguishable subhalos of a single mass $M$; subsequently we show how our expressions generalize for a spectrum of subhalo masses.

The likelihood to observe the measured data for abundance of subhalos in a narrow mass bin, $n = dn/dM \times dM$, is given by 
\begin{align}
P({\rm data}|n) =& P({\rm data}|\mbox{0 subhalos}) P(\mbox{0 subhalos}|n) \nonumber \\
& + P({\rm data}|\mbox{1 subhalo}) P(\mbox{1 subhalo}|n)\nonumber \\
& + P({\rm data}|\mbox{2 subhalos}) P(\mbox{2 subhalos}|n) \nonumber \\
& + \ldots
\label{eqn:decomp}
\end{align}
Each term in this sum represents an integral over the relevant parameter space for 0 subhalos, 1 subhalo, and so on.  The 0 subhalo term is simply the marginalized posterior of our smooth model.  The 1 subhalo term is
\begin{multline}
P({\rm data}|\mbox{1 subhalo}) P(\mbox{1 subhalo}|n) \\
= \int P({\rm data}|x,y) \frac{dP}{dA}(x,y | n) dx dy.
\end{multline}
The two factors in the integrand on the right hand side are given by the marginalized posterior and by Poisson statistics, respectively.  The first factor, representing the marginalized posterior for a subhalo at position $x,y$, is $P({\rm data}|x,y)=P({\rm data}|\mbox{0 subhalos}) \exp(-\Delta\LOGP(x,y)/2)$.  The second factor $\frac{dP}{dA}(x,y|n) dxdy$ represents the Poisson probability to find 1 subhalo within area $dx\,dy$ at position $x,y$, and is independent of $x,y$ because we assume Poisson statistics with uniform mean density.  Indeed, the Poisson probability of finding 1 subhalo at location $x,y$, given mean number density $n$, is $P(1|n) = \exp(-\nu) \nu$, where the expected number $\nu =  n\, dx dy$.  The Poisson likelihood to observe 0 subhalos instead is $P(0|n) = \exp(-\nu)$, so that $P(1|n)=P(0|n) n\,dxdy$. Therefore, we see that  
\begin{multline}
P({\rm data}|1) P(1|n) = P({\rm data}|0) P(0|n) \\ 
\times n \int \exp\left(-\frac{1}{2}\Delta\LOGP(\bm{x})\right) d\bm{x}, 
\end{multline}
where $\Delta\LOGP(\bm{x})$ is given by Equation~\eqref{dchi2}, and represents the improvement in log-posterior between a model with a subhalo at location $\bm{x}$ and a model with no subhalos.  

To compute the 2-subhalo, 3-subhalo, and higher terms in Equation~\eqref{eqn:decomp}, we need analogous $\Delta\LOGP$ maps for all possible configurations and masses of 2 subhalos, 3 subhalos, and so on.  This is an onerous calculation, and typically such integrals are performed using Monte Carlo.  However, since our 1-subhalo maps do not reveal any significant detections (by construction), we can use the 1-subhalo maps to approximate the higher order terms.  Specifically, we assume that each subhalo makes only a perturbative correction to the fit, i.e., $\Delta\LOGP \ll \LOGP_0$, so that the $\Delta\LOGP$ from each subhalo adds linearly to the $\Delta\LOGP$ from any other subhalos.  For example, we assume 
\begin{equation}
\Delta\LOGP(M_1,\bm{x}_1; M_2, \bm{x}_2) \approx \Delta\LOGP(M_1,\bm{x}_1) + \Delta\LOGP(M_2,\bm{x}_2).
\end{equation}
This assumption of linear addition means that we can approximate the $N$-subhalo term in Equation~\eqref{eqn:decomp} using an $N$-fold integral of our 1-subhalo $\LOGP$ maps.  More precisely, the assumption that each subhalo just adds linearly to $\LOGP$ means that every pixel in our $\LOGP$ map is assumed to be independent of every other pixel, which implies that we can construct the total likelihood for number density $n$ by multiplying together the individual constraints on $n$ from all the separate pixels.  For a pixel of area $dA$ with $\Delta\LOGP$ from one subhalo, the constraint on $n$ is
\begin{align}
P(n) &\propto e^{-n\,dA} \left[1 + n\,dA\, e^{-\Delta\LOGP/2} + \ldots\right] \nonumber \\
&= e^{-n\,dA} \sum_N\left[\left(n\,dA\, e^{-\Delta\LOGP/2}\right)^N /\, N! \right] \nonumber \\
&= \exp\left[n\,dA \left(e^{-\Delta\LOGP/2}-1\right) \right].
\end{align}
Multiplying all the pixels, we obtain
\begin{equation}
P(n) = P_0\,\exp\left[n\,dA \sum_i\left(e^{-\Delta\LOGP_i/2}-1\right) \right].
\label{bound}
\end{equation}

We use Equation~\eqref{bound} to determine the constraints on the subhalo abundance from non-detections of subhalos in our $\Delta\LOGP$ maps. 
The assumption of linearity underlying this equation clearly becomes invalid in the limit of large numbers of subhalos that in combination can produce $\Delta\LOGP\,\sim\,\LOGP_0$, or where subhalos overlap with each other.  
Because our 1-subhalo maps do not reveal any significant detections of subhalos, such configurations are probabilistically disfavored, and hence should not lead to significant errors in our approximation. In cases where subhalos do overlap, our tests indicate that our assumption of linear addition tends to underestimate the decrease in posterior probability density, suggesting that our bounds below are (slightly) conservative.  

So far, our discussion has focused on constraining the number density of identical subhalos in a narrow mass bin $dM$.  It is straightforward to generalize Equation~\eqref{bound} to allow for distinguishable subhalos of different masses.  Repeating the same argument used to derive Equation~\eqref{bound} when there are subhalos of varying masses, it is straightforward to see that the likelihood for a mass function $dn/dM$ factorizes into a product of terms from each mass bin,
\begin{align}
P\left(\frac{dn}{dM}\right)&=P_0 \prod_j \exp\left[n_j\,dA \sum_i\left(e^{-\Delta\LOGP_{ji}/2}-1\right) \right] \label{bound1} \\
&= P_0 \exp\left[\sum_{i,j} \frac{dn}{dM}(M_j) dM\,dA \left(e^{-\Delta\LOGP_{ji}/2}-1\right)   \right] \nonumber 
\end{align}
where $i$ runs over angular pixels on the sky, $j$ runs over subhalo mass bins, and we have written the number density of subhalos in bin $j$ as $n_j=dn/dM(M_j)\,dM$.

To summarize, Equation~\eqref{bound1} tells us the effective area over which subhalos are excluded.  Our main lens model tells us the number of subhalos that were detected, and the area over which they were found to occur.  Combining those two measurements, we derive Poisson constraints on the underlying subhalo abundance.

\begin{figure}
\begin{center}
\centering
\includegraphics[trim= 21 0 0 0, clip, width=0.5\textwidth]{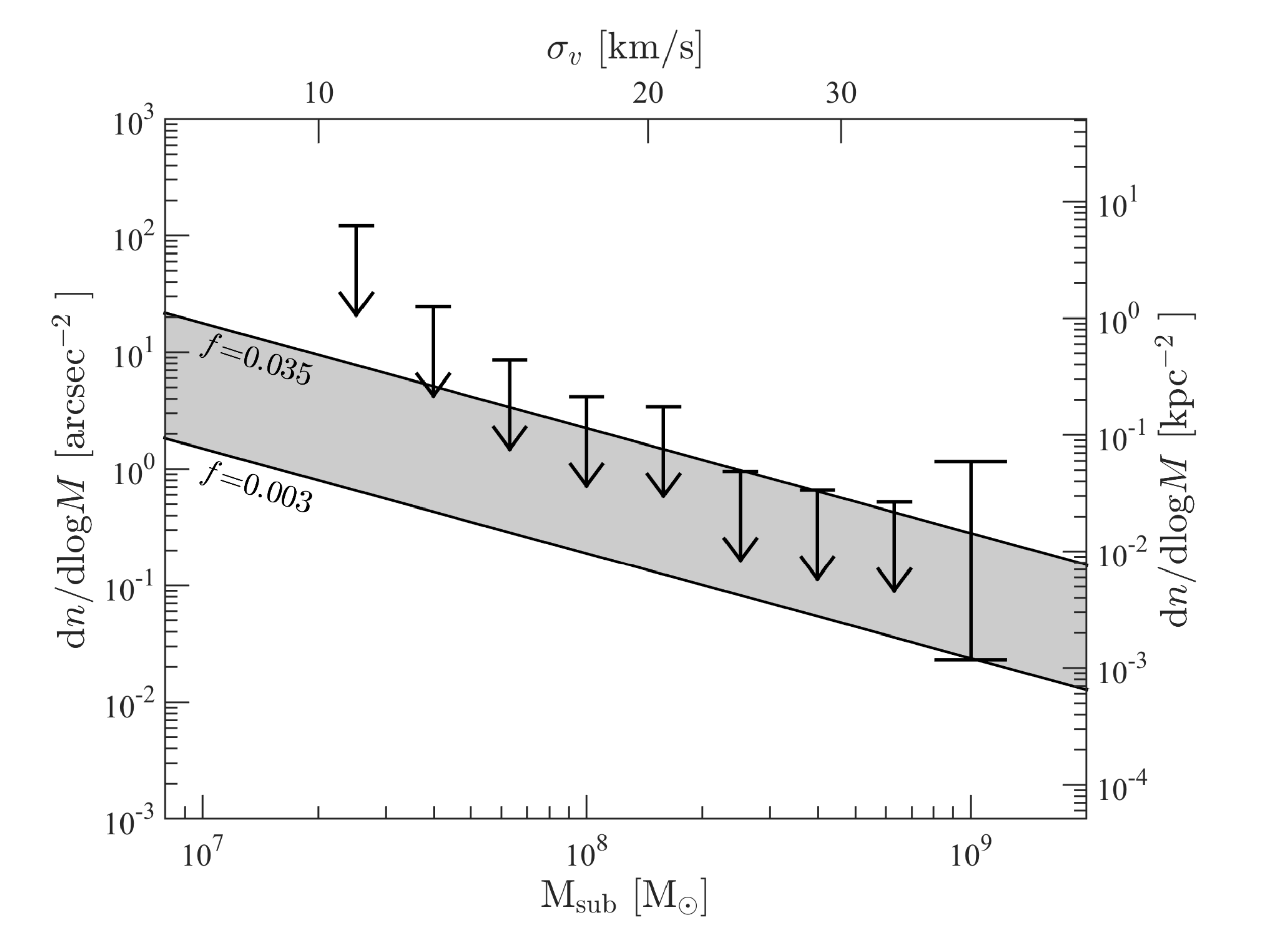}
\centering
\end{center}
\caption{The errorbars indicate the 95\% confidence limits on the projected differential number density of subhalos around SDP.81, derived using the non-detection regions shown in Figure~\ref{fig:subhalo_chi2_maps} and the detection of the $10^9$ M$_{\odot}$ subhalo. For comparison, the shaded band shows the 90\% confidence region from \citet{dalal:02}. 
\label{fig:subhalo_95_limits}}
\end{figure}

Figure~\ref{fig:subhalo_95_limits} shows the resulting constraints on the differential subhalo mass function, $dn/d\log M(M_{\rm sub})$, derived from the maps of $\Delta\LOGP$ shown in Figure~\ref{fig:subhalo_chi2_maps}.  In mass bins where no subhalos were detected, the downward arrows indicate 95\% upper limits.  For the mass bin at $M_{\rm sub}= 10^{9} M_\odot$ where we have a detected subhalo, the central 95\% confidence region is $0.012\ {\rm arcsec}^{-2} < n < 0.2\ {\rm arcsec}^{-2}$. 
If we instead define the confidence region in terms of levels of equal posterior encompassing 95\% of the posterior, we obtain $0.003\ {\rm arcsec}^{-2} < n < 0.1806\ {\rm arcsec}^{-2}$. The reason these two ranges are somewhat different is that the likelihood is asymmetric. 

\begin{figure}
\begin{center}
\centering
\includegraphics[trim= 34 0 0 0, clip, width=0.52\textwidth]{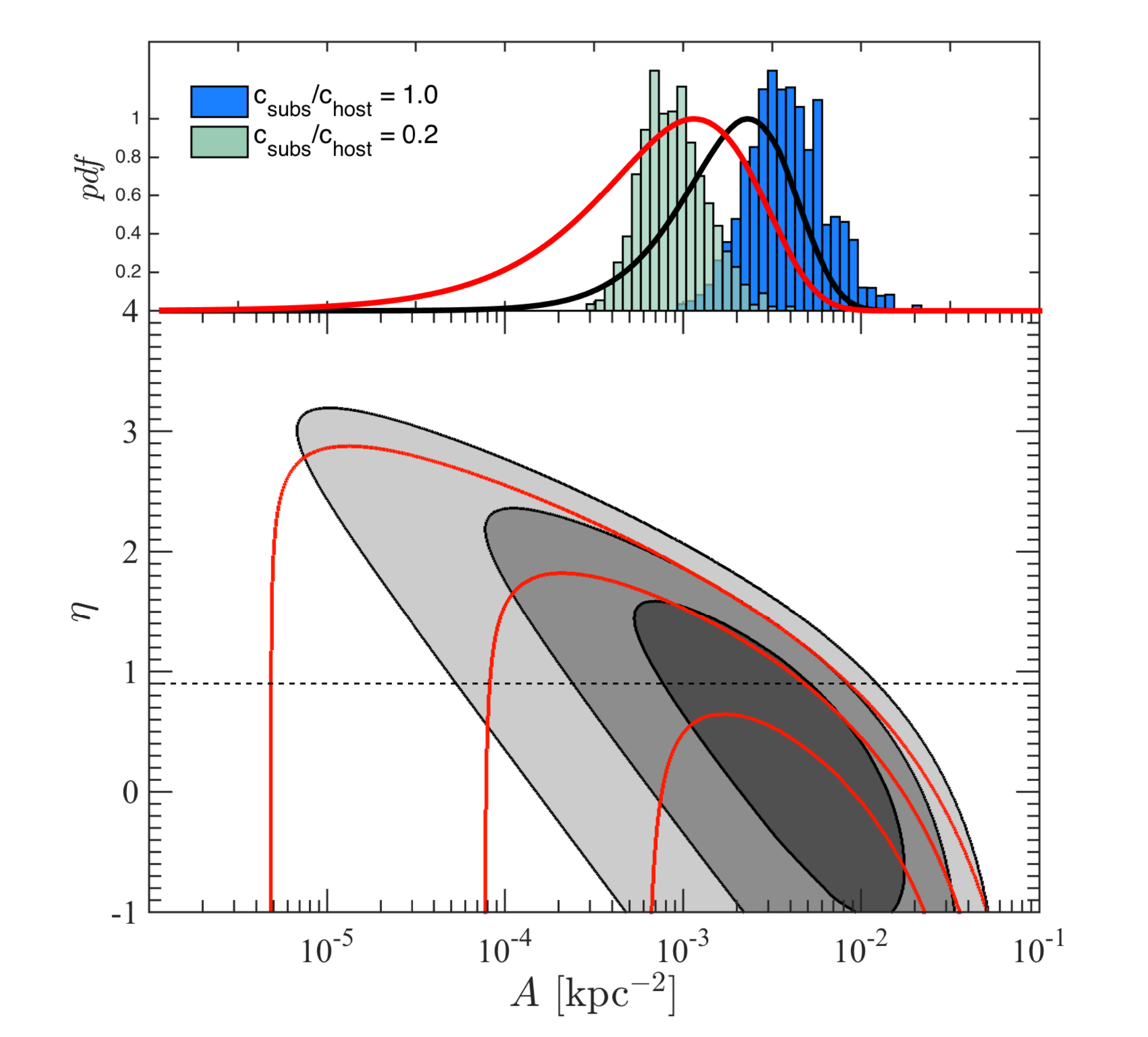}
\centering
\end{center}
\caption{Limits  on  the normalization (A) and slope ($\eta$) of the mass function $dn/d\log M = A (M/M_{\rm pivot})^{-\eta}$, using the bounds in Figure~\ref{fig:subhalo_95_limits}. Here we use $M_{\rm pivot}=10^9 M_\odot$.  The grey contours show constraints derived using Equation~\eqref{bound1}, while the red contours show how the constraints change if we neglect the marginally detected subhalo with $M\approx 10^8 M_\odot$. The top panel shows the probability at $\eta=0.9$. The red and black curves simply show a slice of the probability of the lower panel at $\eta=0.9$. For comparison, the histograms show the distribution of $A$ using assumptions based on $\Lambda$CDM simulations assuming two different values of $c_\text{subs}/c_\text{host}$, which are intended to be representative. These values assume  $\eta=0.9$  and a distribution of host halo masses and concentrations given by abundance matching.  See Section~\ref{sec:lcdm} for details.
\label{fig:SDP81MassFunction}}
\end{figure}

Combining the bounds from the different mass bins, we can derive constraints on the subhalo mass function, using Equation~\eqref{bound1}.  We describe the  mass function using a simple parametrization, $dn/d\log M = A\,(M/M_{\rm pivot})^{-\eta}$, and show in Figure~\ref{fig:SDP81MassFunction} the constraints on these parameters.  
In the next section we compare these constraints to the amount of substructure expected for lens galaxies like SDP.81 in $\Lambda$CDM cosmologies.

\section{Comparison to \texorpdfstring{$\Lambda$}{L}CDM Predictions}
\label{sec:lcdm}

In this section, we compare the constraints on the subhalo abundance in SDP.81 found above, with predictions from $\Lambda$CDM simulations, and also discuss the neighboring environment of this system. To predict the subhalo mass function down to the small masses probed while fully accounting for the halo-to-halo scatter, we follow the methodology presented in \citet{Mao:15},  which captures the dominant source of the halo-to-halo scatter by considering both mass and concentration of host halos. The model is able to reproduce the subhalo abundance found in high-resolution zoom-in simulations~\citep[e.g.,][]{Xu:15} as well as larger statistical samples of halos.

We assume the cumulative subhalo mass function has the form of 
\begin{equation}
\langle n (>M_\text{sub}) \rangle = \left(\frac{M_\text{sub}}{M_0}\right)^{-\eta} - \left(\frac{M_\text{host}}{M_0}\right)^{-\eta},
\end{equation}
where $M_\text{host}$ is the host halo mass, $M_0$ and $\eta$ are the normalization and the log--log slope, respectively, of the subhalo mass function.  We then use $\Lambda$CDM simulations to calibrate the relation between the parameter $M_0$ and the mass and concentration of the host halo. To calibrate this relation, we use the same set of high-resolution zoom-in simulations described in \cite{Mao:15}
with the addition of a very high-resolution cosmological box, ($4096^3$ particles in a $400~\text{Mpc}/h$ box, {\tt ds14\_i}) from the Dark Sky Simulations~\citep{Skillman:14}\footnote{{\tt http://darksky.slac.stanford.edu}}.
This calibration is done by first assuming a constant log--log slope ($\eta$),  then finding the best-fit $M_0$ for each host halo in the simulations, and finally for all host halos, finding the best-fit values of $(\alpha, \beta, \gamma)$ in
\begin{equation}
M_0 = \alpha M_\text{host}^\beta c_\text{host}^\gamma.
\end{equation}
With this model, we can then predict the subhalo mass function given 
the host halo mass and concentration and the log--log slope. 

The subhalo abundance predicted in the procedure described above is for all subhalos within the virial radius of the host halo.
To convert our prediction to the relevant quantity probed by strong lensing measurements, we need to assume a spatial distribution for the subhalos. Here we make three simplifying assumptions: 
(1) the subhalo spatial distribution is \emph{independent} from 
the subhalo mass function (i.e., subhalos of different mass halos have the same spatial distribution);
(2) the angular distribution of subhalos is isotropic \citep[see, however,][]{Nierenberg:11a}; 
and (3) the radial distribution of subhalos within their host halos follows an NFW profile with a characteristic concentration $c_\text{subs}$. In other words, we assume the subhalo abundance factorizes into a mass dependence and radial dependence, $n(M,r) = n(M)\,f(r)$, where the radial dependence $f(r)$ is an NFW profile of concentration $c_\text{subs}$.

To predict the projected abundance of substructure, our model requires a prescription for the concentration of the subhalo distribution, $c_{\rm subs}$.  In $\Lambda$CDM simulations, generally the radial distribution of subhalos is less centrally concentrated than the dark matter distribution of the host halo (i.e., $c_\text{subs}/c_\text{host} < 1$)~\citep[e.g.,][]{Nagai:05,Phoenix}, and at small radii the subhalo distribution may become shallower than an NFW profile~\citep[e.g.,][]{Xu:15}. Observational results for real galaxies are less clear: some are consistent with  $c_\text{subs}/c_\text{host} \simeq 1$ \citep[e.g.,][]{Guo:12,Yniguez:14}, while others imply that galaxies are less concentrated~\citep[e.g.,][]{Hansen:05} than the total mass distribution in their hosts. Also note that our assumption of spherical symmetry might lead us to underestimate the average substructure abundance around lenses, since strong lenses are preferentially viewed along the major axis of their host halos \citep{rozo:07,hennawi:07}. 

Given the uncertainty in predictions for $c_\text{subs}$, we treat it as a free parameter, along with other parameters describing the lens halo: the host halo mass and concentration ($M_\text{host}$, $c_\text{host}$), and the log--log slope ($\eta$) of the subhalo mass function.  Using these model ingredients, we can predict $dn/d\log M$ projected at the Einstein radius.  
The histograms in the top panel of Figure~\ref{fig:SDP81MassFunction} show an example, the distribution of $A$, i.e., $dn/d\log M$ at $M=10^9 M_\odot$ computed with this model.  For this figure, we assume the mass function slope is $\eta=0.9$, and we show two possible values for the subhalo concentration, $c_\text{subs}/c_\text{host}=0.2$ and 1.0, which should span the range of uncertainty described above.
For the other two parameters,
we marginalize over possible values of the host halo mass and concentration using the following prior.  
We first assign galaxy luminosity to dark matter halos and subhalos with the abundance matching technique~\citep[e.g.,][]{Conroy:06,Reddick:13}, and find the joint distribution of mass and concentration corresponding to the luminosity of the lens galaxy, which is $M_r = -21.88 \pm 0.015$ using the SDSS DR10 magnitudes, $k$-corrected to $z=0.1$ using a red galaxy template. 
For the abundance matching procedure, we assume the AGES luminosity function~\citep{Kochanek:12}, and use the maximal circular velocity ($v_{\rm max}$) of the halo at its peak value along its trajectory as the matching proxy, and also apply a constant scatter of 0.2 dex on the luminosity. This gives us the allowed spread of halo mass and concentration typical for galaxies of luminosity similar to the lensing galaxy in SDP.81.  The typical halo mass we found with this procedure is roughly $ 8.7 \times 10^{12} M_\odot h^{-1}$, with an approximate uncertainty of a factor of 3.5. 

\begin{figure*} 
\centering
\includegraphics[trim= 0 0 0 0, clip, width=1.00\textwidth]{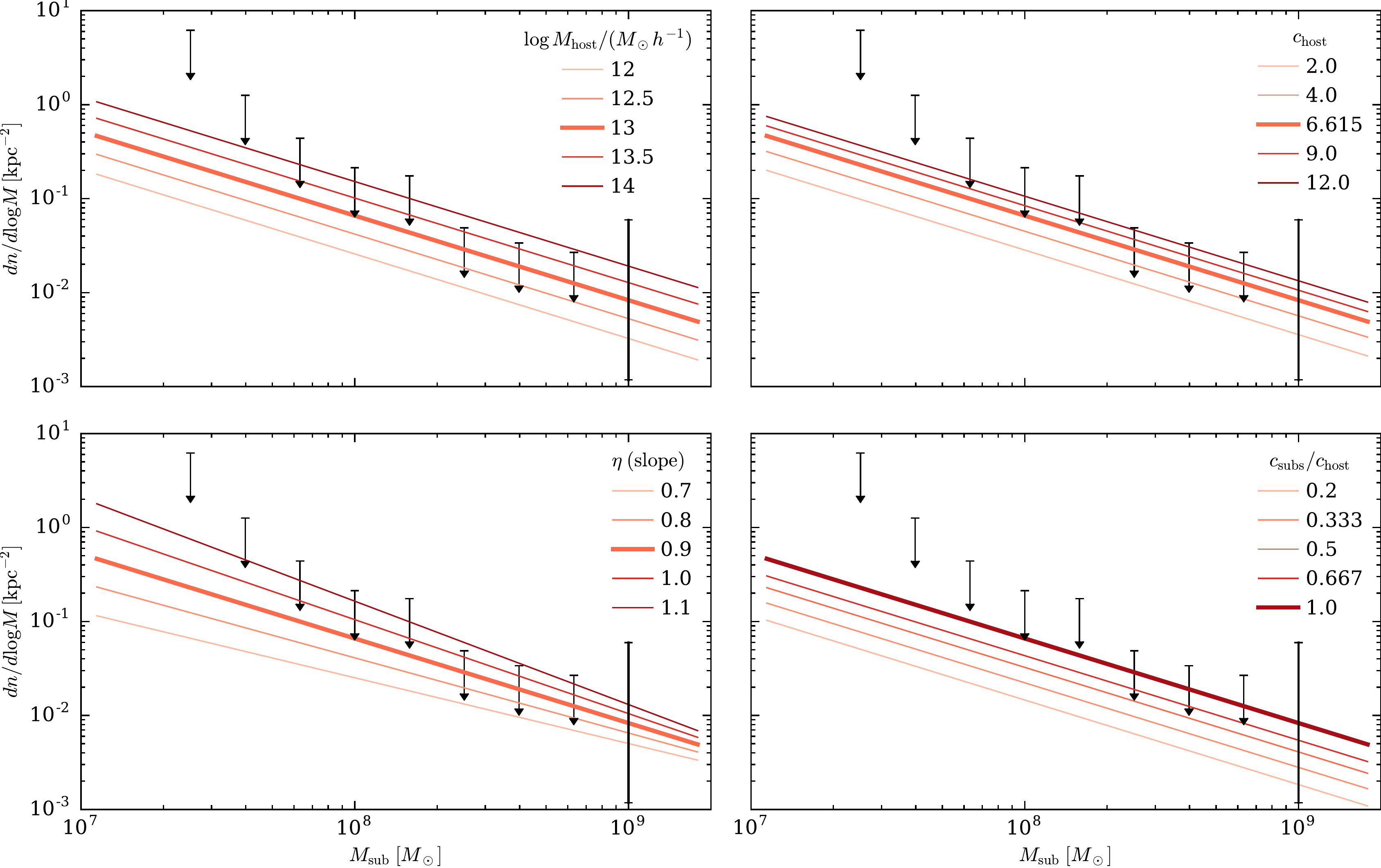}
\caption{ Comparison between the 95\% confidence limits (black squares) 
  on the projected density of substructures, $dn/d\log M$,
  and the predictions (lines) for various properties of the host halo 
  and its substructure distribution.  
  The thick line in each panel assumes  $M_\text{host} = 10^{13} M_\odot h^{-1}$, 
  a median halo concentration for that mass ($c_\text{host}=6.615$), 
  a typical log--log slope ($\eta=0.9$) for the subhalo mass function, 
  and that the radial distribution of subhalos matches that of dark matter
  ($c_\text{subs}/c_\text{host}=1$).   Each panel varies only \emph{one} of the four assumptions listed above.  The varying assumption is showed on the upper right corner of each panel.} 
\label{fig:theory_all}
\end{figure*}

Comparing the lines and the histograms in the top panel of Figure~\ref{fig:SDP81MassFunction}, we can see that substructure limits from SDP.81 are currently consistent with theoretical predictions. The result does hint at a lower normalization of $dn/d\log M$, or equivalently, a smaller value of $c_\text{subs}/c_\text{host}$.
However, the four parameters used here are highly degenerate, because all of them affect the normalization of the projected central density in similar fashions.   
Figure~\ref{fig:theory_all} illustrates this degeneracy by varying only \emph{one} of the four parameters in the model at a time, and shows how each of these four parameters affects the predicted projected density over reasonable ranges in parameters space. 
Given this degeneracy, at this stage it is difficult to jointly constrain these parameters from the observed limits. 
The comparison shown in Figure~\ref{fig:theory_all} also demonstrates, however, that the bounds from this Science Verification dataset are already in an interesting regime that can rule out some portion of the parameter space.
We also note that our estimates for substructure around strong lensing galaxies appear consistent with independent estimates from other high resolution simulations (Fiacconi et al., in prep.). 

The environment of the host halo can also affect the subhalo population.
Fortuitously, SDP.81 is in the SDSS footprint, allowing us to examine its immediate environment.  This system appears to be in close proximity to a massive galaxy cluster in the SDSS DR8 RedMaPPer cluster catalog~\citep{Rykoff:14}.  Here we use RedMaPPer v6.3  (Rykoff et al, in preparation). In the RedMaPPer catalog, this galaxy is a member (with 87\% probability) of a more massive system, with richness $\lambda$ = 32, where $\lambda$ corresponds to the number of red sequence galaxies brighter than $0.4~L_\star$.  Assuming the mass--richness relationship of \cite{Rykoff:14}, this corresponds to a mass of $\sim 2.6\times 10^{14} M_{\odot}$. According to the cluster catalog, SDP.81 is a member galaxy of this system, $640~\text{kpc}~h^{-1}$ away from the most likely central galaxy (in projection). 

The proximity of SDP.81 to a galaxy cluster has the potential to cloud our predictions for its abundance of substructure.  In principle, small subhalos unbound to SDP.81 but within the cluster could project into the strong lensing region.   For the current configuration, we have calculated the expected number of subhalos from such a cluster, at the measured distance (using the same model described in the previous section), and find that these subhalos are subdominant to those expected from the halo of SDP.81.  However, this underscores the importance of determining reliable estimates of lens halo masses and nearby environments before deriving any bounds on cosmological models from observed lensing systems.  

\section{Discussion}
\label{sec:discuss}

In this paper, we present  a method to analyze interferometric measurements of strong gravitational lenses to constrain the abundance of dark matter substructure.  We apply this method to ALMA Science Verification observations of the lens system SDP.81, and we report the detection of a subhalo with mass $M_{\rm sub}\approx 10^9 M_\odot$.  We also find hints for additional substructure at lower mass, but defer a detailed analysis of additional substructure for future work.  We compare our measurements of substructure abundance to previous measurements and to theoretical predictions from CDM, and find that our results are consistent with both.

Although the subhalo analysis presented in this paper only used the continuum data, the method is directly applicable to observations of molecular lines. For example, in this work our joint analysis of bands 6 and 7 treats the two bands as two distinct frequency channels. Analysis of molecular lines, therefore, only differ in the larger number of frequency channels used.
The computational cost of such an analysis, however would be higher. Given the strong hints of lower mass substructure that we find in the continuum data, an analysis of the line data would appear to be quite worthwhile.

The simulations and the analysis of independently generated mock data show that our pipeline can successfully quantify the lensing effects of  $\gtrsim  10^7 M_{\odot}$ subhalos. This framework is designed to be able to marginalize over many nuisance parameters to avoid false detections. We have studied the effects of complex source morphologies, source priors, visibility binning, antenna phase errors, pixel sizes, and grid width, finding that the current pipeline is a reliable tool to quantify the lensing effects of these subhalos, with systematic errors that are below our current statistical uncertainties. 
As we push to lower subhalo masses and lower significance detection or non-detection regimes, however, we will need to quantify more subtle systematic effects. In particular, more detailed studies of weaker interferometric data corruption effects  should be carefully studied. The effects of decorrelation, visibility smearing due to frequency averaging,  choice of parameterization for antenna phase error corrections, and antenna amplitude errors are among these effects.

As Figures~\ref{fig:subhalo_chi2_maps} and \ref{fig:subhalo_95_limits} illustrate, the SDP.81 data set loses sensitivity for detecting individual subhalos at masses $\lesssim 10^7 M_\odot$. Large numbers of subhalos at low masses are expected in all CDM cosmologies, and the deflections from this population of objects combine to form an effectively stochastic field.  Even though the objects generating this random field cannot be individually detected, the collective effects of the population may be significantly detected.  For example, \citet{hezaveh:14a} showed that the power spectrum of density fluctuations of low-mass substructure may be accurately measured using observations of lenses like SDP.81, allowing us to probe the subhalo mass function below our nominal detection limits for individual subhalos.  

Recently, \citet{Inoue2015} published an independent substructure lensing analysis of SDP.81, and it is interesting to compare their conclusions to ours.  \citeauthor{Inoue2015} analyze CLEAN images from continuum band 7, as well as CO 8-7 line data from band 6, and report evidence for substructure at a similar location to our reported detection.  However, the properties of the substructure reported by \citeauthor{Inoue2015} are significantly different than what we find.  In particular, it appears that their results require the presence of a compact, highly underdense region next to one of the arcs.  Underdensities are rarely compact in standard cosmologies, so this result appears puzzling.  Our analysis does not confirm this finding: we find that compact regions of overdensity explain the SDP.81 data far better than adding compact regions of underdensity (or negative density).  One possible origin for the discrepancy, as noted in Section~\ref{sec:implement}, is that analyses of CLEANed images are subject to systematic biases arising from phase errors, which can lead to a host of spurious artifacts in substructure reconstructions (see Figure~\ref{fig:Mars}).  Given the subtlety of the lensing effects of low-mass subhalos, we recommend that substructure analyses operate on the visibilities, and thereby fully extract the information encoded in the interferometric measurements.

In summary, the Science Verification observations of SDP.81 demonstrate the potential of ALMA for probing dark matter structures. Our joint analysis of the bands 6 and 7 data detects a  $M=10^{8.96\pm 0.12} M_{\odot}$ subhalo with a significance of $6.9\sigma$ and produces substructure bounds that are consistent with previous lensing measurements of other systems \citep{dalal:02,vegetti:14}, and also consistent with theoretical expectations as described in Section~\ref{sec:lcdm}. However, the constraints from this one single lens are already interesting, near the abundances expected for halos of this mass.  More importantly, the analysis shows that ALMA data are sensitive to low mass substructure, in a regime that can constrain the properties of dark matter models.  Larger samples of similar lenses have the potential to put tight constraints on the mass function of dark matter substructures. 
Fortunately, large samples of similar lenses are already known from existing submillimeter surveys, suggesting that future ALMA observations have the potential to significantly advance our understanding of the abundance of dark matter substructure.

\acknowledgements{
We are grateful to the \emph{Elemental} team, Jack Poulson and Jeff Hammond,  for their continued help and tremendous support throughout this project. We thank Eli Rykoff for helpful discussions about the environment of SDP.81 in relation to the SDSS ReDMaPPer cluster sample, and Piero Madau for helpful discussions of substructure abundance in simulations.  
Support for this work was provided by NASA through Hubble Fellowship grant HST-HF2-51358.001-A  awarded by the Space Telescope Science Institute, which is operated by the Association of Universities for Research in Astronomy, Inc., for NASA, under contract NAS 5-26555 and by the NSF under Grant No. AST-1212195.
ND is supported by NASA under grant NNX12AD02G, by a Sloan Fellowship, by the Institute for Advanced Study, by the Ambrose Monell Foundation, and by the Center for Advanced Study at UIUC. DM and JV are supported by the U.S. National Science Foundation under award AST-1312950. JEC is supported by NSF awards PLR-1248097 and PHY-0114422. RHW and PJM received support from the U.S. Department of Energy under contract number DE-AC02- 76SF00515. YYM is supported by a Weiland Family Stanford Graduate Fellowship. 
 This research used one of the Dark Sky Simulations, which was run under the INCITE 2014 program using resources of the Oak Ridge Leadership Computing Facility at Oak Ridge National Laboratory,  supported by the Office of Science of the Department of Energy under Contract DE-AC05-00OR22725. Calculations presented in this work were performed on computational facilities provided by the Extreme Science and Engineering Discovery Environment (XSEDE), which is supported by National Science Foundation grant number ACI-1053575, as well as the Blue Waters sustained-petascale computing project, which is supported by the National Science Foundation (awards OCI-0725070 and ACI-1238993) and the state of Illinois. Blue Waters is a joint effort of the University of Illinois at Urbana-Champaign and its National Center for Supercomputing Applications.
This paper makes use of the following ALMA data: ADS/JAO.ALMA\#2011.0.00016.SV. 
ALMA is a partnership of ESO (representing its member states), NSF (USA) and 
NINS (Japan), together with NRC (Canada) and NSC and ASIAA (Taiwan), and KASI (Republic of Korea), in cooperation with the Republic of Chile. The Joint ALMA Observatory is operated by ESO, AUI/NRAO and NAOJ. The National Radio Astronomy Observatory is a facility of the National Science Foundation operated under cooperative agreement by Associated Universities, Inc.}

\bibliographystyle{yahapj}
\bibliography{references}

\end{document}